\documentclass[aps,prd,10pt,showpacs,amsmath,twocolumn,floatfix,amssymb,nofootinbib,longbibliography]{revtex4-2}
\usepackage{graphicx}

\usepackage[normalem]{ulem}
\usepackage[dvipsnames]{xcolor}
\usepackage[utf8]{inputenc}
\usepackage{hyperref}
\hypersetup{
  pdfnewwindow=true,      
  colorlinks=true,        
  linkcolor=PineGreen,    
  citecolor=PineGreen,    
  filecolor=PineGreen,    
  urlcolor=PineGreen      
}
\newcommand{\be}{\begin{eqnarray}}
\newcommand{\ee}{\end{eqnarray}}

\usepackage{marginnote}

\begin{document}

\title{Imaginary time evolution of a quantum system through analytic continuation from real-time quantum simulation}

\author{Peng~Guo}
\email{peng.guo@dsu.edu}
\affiliation{College of Arts and Sciences,  Dakota State University, Madison, SD 57042, USA}

\author{Joshua~Lin}
\email{joshua.lin@anl.gov}
\affiliation{Physics Division, Argonne National Laboratory, Lemont, Illinois 60439, USA}

\author{Anto~Shibu}
\email{anto.shibu@trojans.dsu.edu}
\affiliation{College of Arts and Sciences,  Dakota State University, Madison, SD 57042, USA}

\author{Yong~Zhao}
\email{yong.zhao@anl.gov}
\affiliation{Physics Division, Argonne National Laboratory, Lemont, Illinois 60439, USA}

\date{\today}

\begin{abstract}
Though quantum computing naturally offers an advantage for simulations of real-time quantum systems, implementing Imaginary-Time Evolution (ITE) is comparatively more difficult.
Nevertheless, quantum implementations of ITE are useful both in cases where classical implementations have associated sign problems, and also as an exact method of preparing eigenstates on quantum computers. 
In this work we present an algorithm to obtain ITE of a generic quantum Hamiltonian by performing analytic continuation of measured real-time correlation functions.
We present simulations and demonstrations on IBM quantum hardware of 1D Fokker-Planck equations for classical diffusion process and imaginary-time evolution of integrated correlation functions in 1D quantum mechanical scattering as  examples to demonstrate the effectiveness of the method. 
\end{abstract}

\maketitle

\section{Introduction}\label{sec:intro}
Despite the success of classical monte-carlo methods at studying imaginary-time evolution (ITE) of quantum systems, certain Hamiltonians introduce sign problems that cause classical calculations to become prohibitively expensive as the system size grows~\cite{Loh:1990zz,Nagata:2021ugx,Troyer:2004ge}. 
Separately, there is inherent difficulty in preparing ground states of generic Hamiltonians for use in Quantum Computing~\cite{Kempe:2004sak}. 
Though the relationship between these two problems is complicated~\cite{Bravyi:2007gix}, both can be tackled on quantum computers with a direct implementation of ITE. 

The difficulty is that unlike real-time evolution $e^{-i \hat{H} t}$ (for a given Hamiltonian $\hat{H}$) where there are straightforward trotterization algorithms, ITE is nonunitary. 
Many algorithms have been developed to overcome this difficulty - with different guarantees on convergence and cost. 
For example, QITE~\cite{Motta:2019yya,Nishi2021} approximates the nonunitary operator with a unitary operator which is learned via local tomography. 
VarQITE~\cite{McArdle2019} projects imaginary time evolution onto a variational manifold. 
Another class of methods embeds the non-unitary matrix into a unitary matrix on a larger system, requiring ancilla qubits and post-selection~\cite{PhysRevA.109.052414,Leadbeater_2024}. 
The cost of each of these algorithms differs depending on the specific Hamiltonian considered.

The goal of the present work is to explore the feasibility of   obtaining imaginary-time evolution of  a system from an analytic continuation of real-time correlation functions. 
The main result derived and utilised is that for a hermitian hamiltonian $\hat{H}$ with a lower bound $c$ on the ground state energy, the nonunitary ITE operator can be expressed as a linear combination of real-time evolution operators:
\begin{equation}\label{eq:intro1}
  e^{- \hat{H} \tau}   
  = \frac{i}{2 \pi}\int_{-\infty}^{+\infty} \mathrm{d} t\  \frac{e^{ic(t + i \tau)}}{t + i \tau} e^{-i \hat{H} t}
\end{equation} 
This correspondence offers the attractive possibility of accessing imaginary-time correlation functions simply by (post data-collection) analysis of real-time correlation functions, which are simple to implement on quantum computers. 

Eq.(\Ref{eq:intro1}) expresses ITE matrix elements $\langle \psi_1 | e^{- \hat{H} \tau} | \psi_0 \rangle$ as an integral over real-time matrix elements $\langle \psi_1 | e^{-i \hat{H} t} | \psi_0 \rangle$ for any particular states $\psi_0,\psi_1$ of interest. 
In practice, the real-time matrix elements can be evaluated for some finite number of times $t_n, n \in \{0,\dots,N_\mathrm{samples}\}$ resulting in systematically controlled approximations of the ITE matrix elements. 

Note that this approach to ITE can also be understood exactly as a spectral filter corresponding to the filter $e^{-E\tau}\Theta(E - c)$ where $\Theta$ is the heaviside-theta function. 
If the goal is eigenstate preparation, then given additional information about the low-lying spectrum, as well as information about the overlap of the trial state onto the eigenstate of interest, more optimal filters can be designed~\cite{Ge:2018tcq,Lin:2020acy,Lin:2020zni}. 
The elegance of the ITE filter is its simplicity - corresponding to pure imaginary time evolution, it only requires a lower bound $c$ on the ground state energy. 
Depending on the system, such a lower bound can either be obtained by classical Monte-Carlo methods, explicitly analytically bounded, or determined with quantum algorithms.

Explicit examples of ITE can often be recast as classical stochastic processes. 
In this work, we demonstrate the method's application to 1D Fokker-Planck equations for classical diffusion process, as well as imaginary-time evolution of integrated correlation function in 1D quantum mechanical scattering problems. 
Numerical tests on $1$ and $2$-qubits are performed on IBM hardware, along with comparisons to exact solutions.

The paper is organized as follows. First, the  description of the analytic continuation technique from real-time evolution to imaginary-time evolution  is given in Sec.~\ref{sec:analycontintro}. The specific applications in 1D diffusion process and scattering in quantum mechanics are presented in   Sec.~\ref{sec:application}. Numerical tests carried out on IBM quantum hardware is discussed in Sec.~\ref{sec:numericsIBM},   followed by summary and outlook in Sec.~\ref{sec:summary}.

\section{Accessing imaginary-time evolution  from real-time evolution via analytic continuation}\label{sec:analycontintro}

In this section, we present the basic technique on how to access the ITE operator $e^{- \hat{H} \tau}$ from the real-time evolution operator $e^{- i \hat{H} t}$ by analytic continuation. 
The main equation can be derived directly, and an analysis on truncation errors is also performed in \Ref{ss1}. 
The calculation can also be understood from a contour-deformation perspective, and details of this alternate derivation are presented in \Ref{ss2}. 

\subsection{Direct Approach}\label{ss1}

Given a Hermitian Hamiltonian $\hat{H}$, as well as a lower bound $c$ on the ground state energy, the main formula:
\begin{equation}\label{eq:sec2main}
  e^{- \hat{H} \tau}   
  = \frac{i}{2 \pi}\int_{-\infty}^{+\infty} \mathrm{d} t\  \frac{e^{ic(t + i \tau)}}{t + i \tau} e^{-i \hat{H} t}
\end{equation} 
can be proven directly.
First, note that in the eigenbasis of $\hat{H}$, the $e^{-i\hat{H}t}$ factor contributes an $e^{-iEt}$ to the integral for the eigenstate with energy $E$. 
The condition $c < E$ allows one to close the $t$-contour in the negative imaginary $t$ half-plane, picking up the residue at $t = -i \tau$. 

Supposing that the lower bound $c$ was chosen poorly, and that there is actually an eigenenergy $E < c$, the resulting integral can be closed in the \textit{upper} half-plane, and evaluates to exactly $0$ (no residue to pick up). 
It is in this sense that the right hand side of Eq.(\Ref{eq:sec2main}) acts as a spectral filter of the form $e^{-E\tau} \Theta(E-c)$. 

In practice, the integral in Eq.(\Ref{eq:sec2main}) is approximated by first truncating the domain of integration to $\int_{-t_\mathrm{cut}}^{t_\mathrm{cut}}\mathrm{d}t$ for some chosen cutoff $t_\mathrm{cut}$, followed by an approximation of the integral by evaluating the integrand at finitely many chosen points $t_n$, $n \in \{0,\dots,N_\mathrm{samples}\}$. 
The relative error due to the truncation to $t_\mathrm{cut}$ can be explicitly bounded, for $t_\mathrm{cut} > \tau$:
\begin{equation}
\left|\frac{\frac{i}{2 \pi} \int_{-t_\mathrm{cut}}^{+t_\mathrm{cut}} \mathrm{d}t\  \frac{e^{ic(t+i\tau)-iEt}}{t+i\tau}}{e^{-E\tau}} - 1\right| \leq \frac{2}{\pi}\frac{t_\mathrm{cut} + \tau}{t_\mathrm{cut}^2 + \tau^2} \frac{e^{(E-c)\tau}}{E-c}
\end{equation}
with details presented in subsection~\Ref{subsec:DAB}. 
For a fixed $c$, increasing $t_\mathrm{cut}$ leads to a $1/t_\mathrm{cut}$ convergence to the correct ITE result. 
Note that choosing $c$ tightly in general improves the convergence (by suppressing the exponential factor in the numerator), however choosing $c$ too close to the ground state energy $E_0$ also leads to a $\frac{1}{E-c}$ blow-up in the relative error, which can also be confirmed numerically. 
In the numeric examples demonstrated in this work, the approximation to the truncated integral is then carried out via gaussian quadrature. 

\subsection{Contour Deformation approach}\label{ss2}

Alternatively, the same result can also be derived by considering a contour deformation in the complex energy plane. 

  \begin{figure}[h]
  \includegraphics[width=0.45\textwidth]{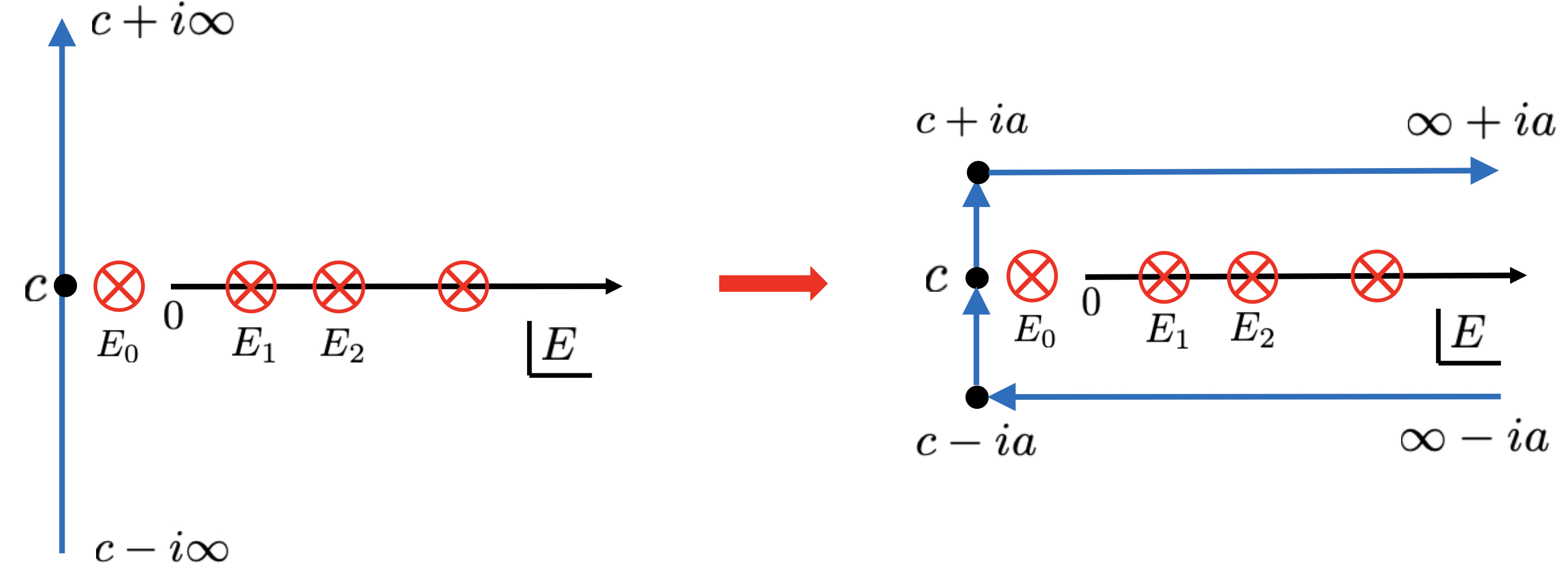}
\caption{The deformation of  contour  $\int_{c- i \infty}^{c + i \infty}  \mathrm{d} E $  into $ \left ( \int_{\infty - i a}^{  c- i a } +  \int_{c- i a}^{c} + \int_c^{c+ia}  + \int_{ c + i a }^{\infty + i a} \right ) \mathrm{d} E $. }
\label{Fig:contourintegral}
\end{figure}

We start with rewriting  $e^{- \hat{H} \tau}$ though a contour integral of Green's function operator $1/(E- \hat{H})$:
\begin{equation}
e^{- \hat{H} \tau} = \frac{1}{2\pi } \int_C  \mathrm{d} E\  \frac{e^{- E \tau} }{E - \hat{H}} , \label{eq:realtoimagintegral}
\end{equation}
where the contour $C$ can be chosen as  a straight line that is parallel to imaginary axis: $\int_C   \mathrm{d} E= \int_{c - i \infty}^{c + i \infty}  \mathrm{d} E $ \cite{Watson:1956zz}. We will also assume that    the eigenvalues of Hamiltonian have a lower bound and the lowest energy eigenvalue is referred as $E_0$.  The singularities of Green's function operator $1/( E - \hat{H} )$ thus are poles and a branch cut sitting on real axis in complex $E$-plane, and the singularities are bound by the lowest pole located at $E_0$.
The straight line of contour $C$ is positioned to sit below $E_0$ by  setting the  real valued  free  parameter $c $ to  $c <E_0$, as shown in the left panel of Fig.~\ref{Fig:contourintegral}. {The contour integral in Eq.~(\ref{eq:realtoimagintegral}) picks up residues of poles or imaginary parts of branch cuts in the Green's function operator, such that
\begin{equation}
\frac{i}{2\pi } \int_C \mathrm{d} E\    \frac{e^{- E \tau} }{E - \hat{H}} = \sum_n e^{- \epsilon_n \tau} | n \rangle \langle n |,  
\end{equation}
where     the $\epsilon_n$ and $| n \rangle$ are discrete eigensolutions of Hamiltonian operator $\hat{H}$: 
\begin{equation}
\hat{H} | n \rangle = \epsilon_n | n \rangle.
\end{equation}
 We remark that the Eq.(\ref{eq:realtoimagintegral}) applies to the continuum spectra as well. 
The Cauchy's integral theorem warrants that all shapes of contour $C$   yield   exactly the same result as far as the singularities of Green's function are enclosed in the closed contour $C$. The contour $C$ can be chosen based on convenience.

Next, we will show that  a connection between Green's function operator $1/(E-\hat{H})$ and $e^{- i \hat{H} t}$ through an integral over $t$ can be established, where we want to make sure the $t$ is defined on positive real axis so that $e^{- i \hat{H} t}$ operator can be evaluated on quantum computer efficiently. 
To make sure the convergence  of the   integral of $ e^{- i \hat{H} t} $ operator over time, we first deform  the contour $C$ running along a straight line into four pieces, see right panel of Fig.~\ref{Fig:contourintegral},
\begin{equation}
\int_{c- i \infty}^{c + i \infty}  \mathrm{d} E = \left ( \int_{\infty - i a}^{  c- i a } +  \int_{c- i a}^{c  } +  \int_{c }^{c + i a}  + \int_{ c + i a }^{\infty + i a} \right ) \mathrm{d} E, \label{eq:contoursplit}
\end{equation}
where $a$ is a positive real number. The double integrals in  Eq.(\ref{eq:realtoimagintegral}) are now split into four pieces. Hence, Eq.(\ref{eq:realtoimagintegral}) is now given by
\begin{align}
&e^{- \hat{H} \tau} \nonumber\\
&= \frac{1}{2\pi  i}   \int_{c }^{  \infty }   \mathrm{d} \epsilon\     \frac{ e^{- (\epsilon +  i a) \tau} }{ \epsilon + i a - \hat{H}}    - \frac{1}{2\pi }    \int_{0 }^{ a}   \mathrm{d} \epsilon \   \frac{e^{- (c+  i \epsilon) \tau} }{c+ i \epsilon - \hat{H}} \nonumber\\
& + \frac{1}{2\pi  i}   \int_{c }^{  \infty }   \mathrm{d} \epsilon\     \frac{ e^{- (\epsilon -  i a) \tau} }{ \epsilon - i a - \hat{H}}    - \frac{1}{2\pi }    \int_{0 }^{ a}   \mathrm{d} \epsilon \   \frac{e^{- (c-  i \epsilon) \tau} }{c- i \epsilon - \hat{H}} , \label{eq:expHtausplit}
\end{align}
after change of variables.
Using the identity
  \begin{equation}
 -i  \int_0^\infty \mathrm{d} t\  e^{i E t} e^{- i \hat{H} t}  = \frac{1}{E - \hat{H}},  \label{eq:realtoGreen}
\end{equation}
where the condition of $\mathrm{Im} [E] > 0$ must be met to make sure the convergence of integration, we can rewrite the  first two terms in Eq.(\ref{eq:expHtausplit})  to,
\begin{align}
 &  \frac{1}{2\pi  i}   \int_{c }^{  \infty }   \mathrm{d} \epsilon\     \frac{ e^{- (\epsilon +  i a) \tau} }{ \epsilon + i a - \hat{H}}    - \frac{1}{2\pi }    \int_{0 }^{ a}   \mathrm{d} \epsilon \   \frac{e^{- (c+  i \epsilon) \tau} }{c+ i \epsilon - \hat{H}} \nonumber \\
 & = \frac{i}{2\pi }    \int_{0 }^{ a}   \mathrm{d} \epsilon \ e^{- (c+i \epsilon) \tau}  \hat{I}_1 (\epsilon)    +  \frac{1}{2\pi }   \int_{c }^{  \infty }   \mathrm{d} \epsilon  \ e^{- (\epsilon +  i a) \tau}  \hat{I}_2 (\epsilon) ,
\end{align}
 where
\begin{align}
\hat{I}_1 (\epsilon) & = \int_0^\infty \mathrm{d} t \ e^{i (c+i \epsilon)  t} e^{- i \hat{H} t} = \frac{i}{c+  i \epsilon - \hat{H}}  , \nonumber \\
\hat{I}_2 (\epsilon) & =  \int_0^\infty \mathrm{d} t \ e^{i (\epsilon + i a) t} e^{- i \hat{H} t} =  \frac{i}{ \epsilon + i a - \hat{H}}    . \label{eq:I12functions}
\end{align}
 Also using relations
\begin{align}
 &   \frac{1}{ \epsilon - i a - \hat{H}}  =\left [  \frac{1}{ \epsilon + i a - \hat{H}}  \right ]^\dag  = i \hat{I}_2^\dag  (\epsilon) ,  \nonumber \\
&  \frac{1}{c- i \epsilon - \hat{H}} = \left [  \frac{1}{c+  i \epsilon - \hat{H}} \right ]^\dag = i \hat{I}^\dag_1 (\epsilon)  ,
\end{align}
the last two terms  in Eq.(\ref{eq:expHtausplit})  can be rewritten  in terms of  complex conjugate of  $\hat{I}_{1,2} (\epsilon)$ functions:
 \begin{align}
 & \frac{1}{2\pi  i}   \int_{c }^{  \infty }   \mathrm{d} \epsilon\     \frac{ e^{- (\epsilon -  i a) \tau} }{ \epsilon - i a - \hat{H}}    - \frac{1}{2\pi }    \int_{0 }^{ a}   \mathrm{d} \epsilon \   \frac{e^{- (c-  i \epsilon) \tau} }{c- i \epsilon - \hat{H}} \nonumber \\
 & = \frac{1}{2\pi  }   \int_{c }^{  \infty }   \mathrm{d} \epsilon\   e^{- (\epsilon -  i a) \tau}   \hat{I}_2^\dag (\epsilon)     - \frac{i}{2\pi }    \int_{0 }^{ a}   \mathrm{d} \epsilon\  e^{- (c-  i \epsilon) \tau}    \hat{I}_1^\dag (\epsilon)  . 
\end{align}

  \begin{figure*}
 \centering
 \includegraphics[width=0.32\textwidth]{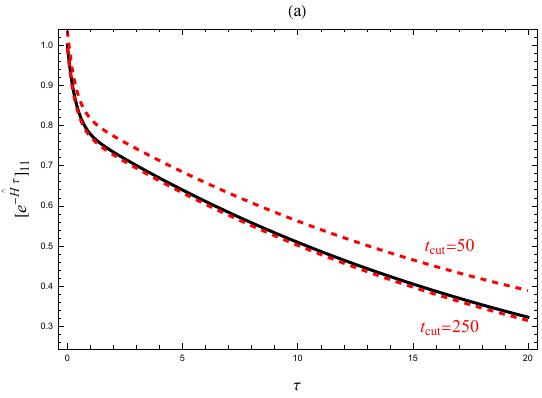}
 \includegraphics[width=0.32\textwidth]{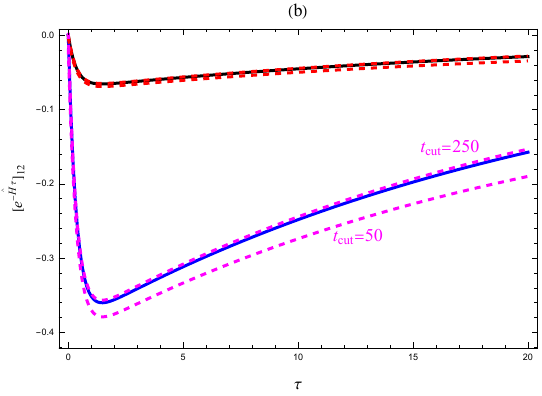}
 \includegraphics[width=0.32\textwidth]{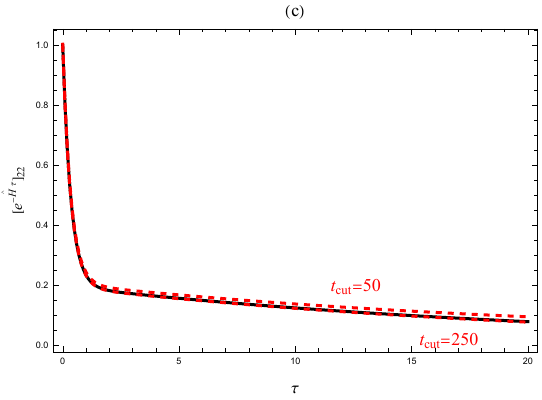}
 \caption{   Comparison of $e^{-\hat{H} \tau}$  (black solid for real part and blue solid for imaginary part) vs the results via analytic continuation from $e^{- i \hat{H} t}$ (red dashed for real part and purple dashed for imaginary part) by Eq.(\ref{eq:mainresult}) with a $2\times 2$ Hamiltonian matrix given in Eq.(\ref{eq:exampleHmatrix}): (a) $\left [ e^{- \hat{H} \tau} \right ]_{11}$, (b) $\left [ e^{- \hat{H} \tau} \right ]_{12} = \left [ e^{- \hat{H} \tau} \right ]^*_{21}$, and (c) $\left [ e^{- \hat{H} \tau} \right ]_{22}$.  
The parameters are taken as:   $E_0 =0.6 $, $g =0.2+ i 1.1 $, $E_1=2.3$,  $ a =1$, and $c=0$. The cutoffs on $t$ integration are taken $t_\mathrm{cut}=50$ and $250$ respectively, with $t_\mathrm{cut} = 50$ showing some systematic error that is largely removed when increased to $t_\mathrm{cut} = 250$.   }\label{Fig:expHtplot}
\end{figure*}

 Now, the imaginary-time evolution operator $e^{- \hat{H} \tau}$ can be computed from  the real-time evolution operator $e^{-  i\hat{H} t}$  by numerically convergent double integrals,
\begin{align}
  e^{- \hat{H} \tau}   
 & =  \frac{i}{2\pi }    \int_{0 }^{ a}   \mathrm{d} \epsilon\   e^{- c \tau}   \left [  e^{- i \epsilon  \tau}  \hat{I}_1 (\epsilon) -  e^{ i \epsilon  \tau}    \hat{I}^\dag_1 (\epsilon) \right ]  \nonumber \\
 & +  \frac{1}{2\pi }   \int_{c }^{  \infty }   \mathrm{d} \epsilon \  e^{- \epsilon   \tau}\left [ e^{-    i a \tau} \hat{I}_2 (\epsilon) + e^{  i a \tau}   \hat{I}^\dag_2 (\epsilon)   \right ] . \label{eq:mainresult}
\end{align}
Explicitly performing the $\epsilon$ integrals in Eq.~\Ref{eq:mainresult} (after factoring out the $\epsilon$ phase dependence of $I_1(\epsilon),I_2(\epsilon)$) causes the $a$-dependence to disappear, and leads to the simplified result shown in Eq.~\Ref{eq:sec2main}. The cutoff dependence caused by truncating the real-time integrals to $\int_{-t_\mathrm{cut}}^{+t_\mathrm{cut}}\mathrm{d}t$ are analysed from the contour-deformation perspective in subsection~\Ref{subsec:CDB}.

\subsection{$2 \times 2$ example}
As an example, the relation in Eq.(\ref{eq:mainresult}) can be demonstrated with a $2\times 2$ hermitian Hamiltonian matrix
\begin{equation}
\hat{H} = \begin{bmatrix} E_0 & g \\ g^* & E_1 \end{bmatrix}, \label{eq:exampleHmatrix}
 \end{equation}
where $E_0,E_1$ are real, $g$ is a complex number. In Fig.~\ref{Fig:expHtplot} we show comparisons between $e^{-\hat{H} \tau}$ and the results obtained by analytic continuation  from $e^{- i \hat{H} t}$ for two different values of the cutoff, $t_\mathrm{cut} = 50, t_\mathrm{cut} = 250$. We also plot  the asymptotic relation
\begin{equation}
\langle x| e^{- \hat{H} \tau} | x \rangle  \stackrel{\tau \rightarrow \infty}{\longrightarrow}  | \langle x | E_- \rangle |^2 e^{  - E_- \tau}
\end{equation}
 in Fig.~\ref{Fig:diagexpHtplot},
where $ x =0,1$, and  $E_-$  and $| E_-  \rangle $ are  the ground state eigen-solutions of $\hat{H}$:
\begin{equation}
E_- =  \frac{E_0 + E_1 - \sqrt{ (E_1-E_0)^2 +4 |g|^2 } }{2} ,
\end{equation}
and
\begin{equation}
| E_-  \rangle   = \frac{1}{ \sqrt{ N_{E_-} } }  \begin{bmatrix}   \frac{ E_0- E_1 - \sqrt{ (E_0 - E_1)^2 + 4 |g|^2 } }{ 2g^*}  \\ 1 \end{bmatrix}     ,
\end{equation}
where $N_{E_-} = 1+ \frac{ \left [  E_0- E_1 - \sqrt{ (E_0 - E_1)^2 + 4 |g|^2 } \right ]^2 }{ 4 |g|^2}  $ is the normalization factor.

  \begin{figure}
\includegraphics[width=0.45\textwidth]{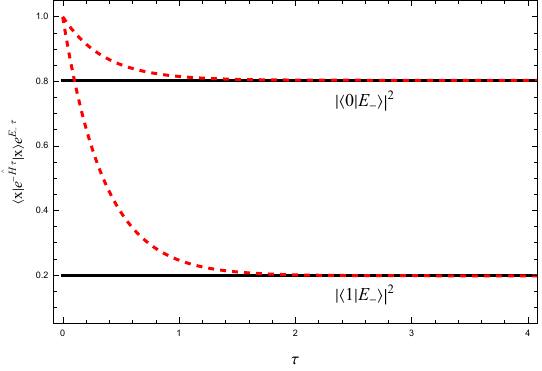}
\caption{  Comparison betweens   $\langle x| e^{- \hat{H} \tau} | x \rangle e^{ E_- \tau}$ via analytic continuation of real-time matrix elements  $\langle x| e^{-  i \hat{H} t} | x \rangle $  (dashed red) vs. exact results of $| \langle x | E_- \rangle |^2 $ (black solid), where $E_-$, $| E_-  \rangle $ are  the ground state energy and eigenvector of $\hat{H}$.  
}\label{Fig:diagexpHtplot}
\end{figure}

\section{Applications in 1D systems}\label{sec:application} 

In this section, we will demonstrate the  feasibility and applicability  of analytic continuation technique with three  exactly solvable   1D examples of ITE: (1) Brownian motion with a constant drift; (2) Ornstein-Uhlenbeck Process; and (3) scattering  in quantum mechanics. Both diffusion processes,  Brownian motion with a constant drift and Ornstein-Uhlenbeck Process, can be described by Fokker-Planck equations with different drift coefficients, one is given by a constant and another is  a linear function. For the sake of convenience,    some technical details and summarized results  of 1D Fokker-Planck equations for  Brownian motion with a constant drift and Ornstein-Uhlenbeck Process are presented  in  Appendix \ref{sec:appendix-FPequation}. The focus in this section will be given to the demonstration of analytic continuation technique in these applications, {and all the numerical results are carried out on classical computers by matrix computation. }

\subsection{1D Fokker-Planck equations}\label{sec:1DFPequation} 
A single-variable Fokker-Planck equation can be converted into a  imaginary time   Schr\"odinger-like equation, see details given in Appendix \ref{sec:appendix-FPequation}. Ultimately we want to solve imaginary time evolution of a  state, 
    \begin{equation}
  | e^{\frac{\Phi}{2}}  W( \tau)  \rangle  = e^{-  \hat{H}  \tau } | e^{\frac{\Phi}{2}} W( 0)  \rangle , \label{eq:Wtstate}
 \end{equation}
 where the projection of state into position is given by
   \begin{equation}
 \langle x | e^{\frac{\Phi}{2}}  W( \tau)  \rangle  = e^{\frac{\Phi (x)}{2}}  W( x, \tau).
 \end{equation}
 The $W(x, \tau)$   is the probability distribution function of a diffusion process that is the solution of Fokker-Planck equation in Eq.(\ref{eq:FPLequation}) and Eq.(\ref{eq:FPLdef}). The $\Phi(x)$ function is related the drift and diffusion coefficients of Fokker-Planck equation, $\gamma (x)  $ and $D$, respectively  through a integral by,
\begin{equation}
\Phi (x) = - \int^x_0 \frac{ \gamma (x')  }{D}d x' .
\end{equation}
The drift coefficients for (1) Brownian motion with a constant drift (BM) and  (2) Ornstein-Uhlenbeck (OU) Process are given by 
\begin{equation}
\gamma (x) = \begin{cases} \gamma , &  \mbox{BM process}; \\ -\gamma x , & \mbox{OU process} \ \  (\gamma >0). \end{cases}
\end{equation}
  The analytic expressions of $\Phi(x)$ function for two processes are
\begin{equation}
\Phi (x) = \begin{cases}  - \frac{ \gamma  }{D}   x, &  \mbox{BM process}; \\  \frac{ \gamma  }{2 D}  x^2, & \mbox{OU process} \ \  (\gamma >0). \end{cases}
\end{equation}
The Hermitian Hamiltonian operator $\hat{H}$ is related to the drift and diffusion coefficients by
 \begin{equation}
  \hat{H}    = -    D \frac{\partial^2}{\partial x^2}   +    \frac{\gamma^2(x)}{4D}   +  \frac{1}{2}   \frac{\partial  \gamma(x) }{\partial x}  ,
\end{equation}
hence
\begin{equation}
 \hat{H}  = \begin{cases} -    D \frac{\partial^2}{\partial x^2}   +   \frac{ \gamma^2}{4 D}    , &  \mbox{BM process}; \\ -    D \frac{\partial^2}{\partial x^2}  - \frac{\gamma}{2}   +   \frac{ \gamma^2  x^2 }{4 D}  , & \mbox{OU process} \ \  (\gamma >0). \end{cases}
\end{equation}
The Hamiltonian operator $\hat{H}$  for Brownian motion with a constant drift  process is equivalent to the Hamiltonian of a free particle in quantum mechanics by a constant energy shift, which yields the continuous eigen-energies  and    plane wave eigen-solutions. The Hamiltonian operator $\hat{H}$  for Ornstein-Uhlenbeck  Process is equivalent to the confining harmonic oscillator Hamiltonian  in quantum mechanics by a constant energy shift, which yields the discrete energy spectra of equal gap and zero valued ground state energy.

  \begin{figure}
\includegraphics[width=0.45\textwidth]{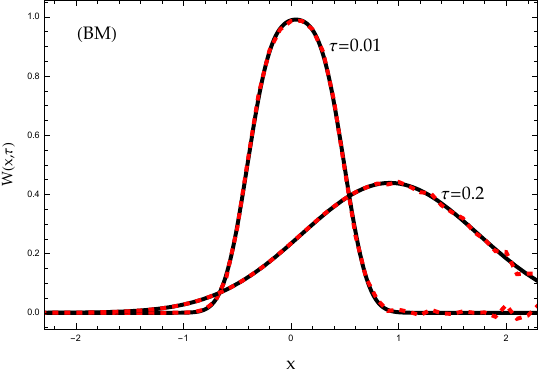}
\includegraphics[width=0.45\textwidth]{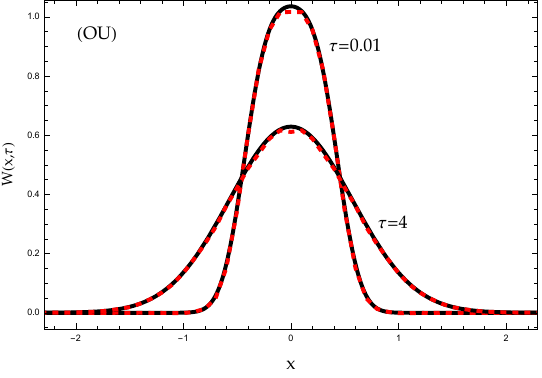}
\caption{ Probability distributions $W(x,\tau)$ for  Brownian motion with a constant drift process (upper panel) and  Ornstein-Uhlenbeck  Process  (lower panel).  The exact solutions (solid black) given by Eq.(\ref{eq:wtexactBM}) for   Brownian motion with a constant drift process  and by Eq.(\ref{eq:wtexactOU})  for Ornstein-Uhlenbeck  Process   compared with  results (red dashed)  by analytical continuation  of   real-time evolution of $ e^{-  i \hat{H}  t } | e^{\frac{\Phi}{2}}  W( 0)  \rangle $ by using Eq.(\ref{eq:mainresult}).   The parameters are: $\gamma = 4.6$, $D = 1.5$, $a_x=0.1$,  $a=1.0$ and $c  =0$ and $  -0.2$ for BM and OU processes respectively. The size of Hamiltonian matrix is $81\times 81$. The cutoff on $t$ integration is taken   $t_{cut}=100$.  The initial condition of $W(x,0)$ is given by Eq.(\ref{eq:w0initial}), where $b=0.9$.  }\label{Fig:wtplot}
\end{figure}

Using the initial condition of a rectangle shape of width $b$ and height one, 
\begin{align}
  W(x  ,0 ) = \begin{cases} 1 , &  x \in [- \frac{b}{2}, \frac{b}{2}] ,  \\ 0 , & \mathrm{otherwise},  \end{cases}  ,    \label{eq:w0initial}
\end{align}
  the analytic solution of probability distribution functions are thus given by Eq.(\ref{eq:Wtauevolution}) together with Eq.(\ref{eq:ProbsolutionBM}) and Eq.(\ref{eq:ProbsolutionOU}),
 \begin{equation}
  W(x, \tau)   =    \frac{\mbox{erf} \left (   \frac{\frac{b}{2} -x + \gamma \tau}{\sqrt{4 D \tau}} \right ) + \mbox{erf} \left (   \frac{\frac{b}{2} + x - \gamma \tau}{\sqrt{4 D \tau}} \right )}{2} , \label{eq:wtexactBM}
\end{equation}
for Brownian motion with a constant drift process and
\begin{align}
  W(x, \tau)   
  & =  \frac{ \mbox{erf}\left ( \frac{ \sqrt{ \gamma} ( \frac{b e^{-\gamma \tau}}{2} - x) }{\sqrt{ 2 D \left (1- e^{-2 \gamma \tau} \right ) }}  \right ) + \mbox{erf}\left ( \frac{ \sqrt{ \gamma} ( \frac{b e^{-\gamma \tau}}{2} + x)}{\sqrt{  2 D \left (1- e^{-2 \gamma \tau} \right ) }}  \right ) }{2 e^{-\gamma \tau}}      \nonumber \\
&    \stackrel{ \tau \rightarrow \infty}{\rightarrow}     b  \sqrt{ \frac{\gamma }{2 \pi D }}  e^{-   \frac{\gamma  x^2 }{2 D  } }   , \label{eq:wtexactOU}
\end{align}
for Ornstein-Uhlenbeck  Process.  {We remark that the initial condition in Eq.(\ref{eq:w0initial}) is not normalized, so it is related to the probability distribution by a constant normalization factor. } The exact solutions    can be compared with the results by analytic continuation using Eq.(\ref{eq:mainresult}) from real-time evolution of $ e^{-  i \hat{H}  t } | e^{\frac{\Phi}{2}}  W( 0)  \rangle $. The  real-time evolution of  the state $ e^{-  i  \hat{H}  t } | e^{\frac{\Phi}{2}}  W( 0)  \rangle $ is solved numerically by discretizing  the coordinate $x = a_x \alpha$ where $\alpha \in \mathbb{Z}$,  where the discretized 
Hamiltonian operator  is given by
\begin{equation}
\hat{H} =  \sum_{\alpha \in \mathbb{Z}}  V(\alpha)  | \alpha \rangle \langle \alpha |      -  \frac{  D }{a_x^2}  \sum_{\alpha \in \mathbb{Z}}  \left (  | \alpha  \rangle \langle \alpha +1  |  + | \alpha  \rangle \langle \alpha -1  | \right ) , \label{eq:FPHmatrix}
\end{equation}
where
\begin{equation}
V (\alpha)  = \begin{cases}  \frac{2 D}{a_x^2}  + \frac{  \gamma^2  }{4D}     , &  \mbox{BM process}; \\   \frac{2 D}{a_x^2}   - \frac{\gamma}{2}   +   \frac{ \gamma^2  a_x^2 \alpha^2 }{4 D}  , & \mbox{OU process} \ \  (\gamma >0). \end{cases} \label{eq:HVFPdef}
\end{equation}
The demo plots of exact solutions of $W(x, \tau)$ given by Eq.(\ref{eq:wtexactBM}) and Eq.(\ref{eq:wtexactOU})  for   Brownian motion with a constant drift process  and   Ornstein-Uhlenbeck  Process respectively vs. the results by    analytic continuation using Eq.(\ref{eq:mainresult}) from real-time evolution of $ e^{-  i \hat{H}  t } | e^{\frac{\Phi}{2}}  W( 0)  \rangle $ are given in Fig.~\ref{Fig:wtplot}. The exact solutions and analytic continuation results show excellent agreements in Fig.~\ref{Fig:wtplot}.

\subsection{Imaginary-time evolution of integrated correlation function in 1D quantum mechanical scattering}\label{sec:1Dctquantum}

Next let's consider an exactly solvable 1D quantum mechanical scattering problem used in Refs.~\cite{Guo:2023ecc,Guo:2024pvt,Guo:2026qkx,Guo:2026nuc}. The  Hamiltonian of  1D quantum mechanical particles  interacting  with a contact interaction potential is given by
\begin{equation}
\hat{H} = - \frac{1}{2m}  \frac{d^2}{d x^2} + V_0 \delta(x),
\end{equation}
where   $V_0$ denotes the strength of contact interaction.   As shown in  Refs.~\cite{Guo:2023ecc,Guo:2024zal,Guo:2024pvt,Guo:2025vgk,Guo:2025ngh,Guo:2025lmd},  the infinite volume  two-particle elastic scattering phase shift, $\delta (\epsilon )$, is  related  to the imaginary time integrated two-particle correlation function   through a weighted integral,
\begin{equation}
C(\tau) - C_0 (\tau)  =  \frac{\tau}{\pi} \int_0^{\infty} d \epsilon \delta(\epsilon) e^{-  \epsilon \tau},  \label{eq:ICFmainEQ}
\end{equation}
 where $C(\tau)$ and $C_0 (\tau)$ refer to the interacting and non-interacting integrated correlation functions respectively.   
 The spectral representation of integrated correlation function  is given by
 \begin{equation}
 C(\tau ) = \mathrm{Tr} \left [ e^{-  \hat{H} \tau} \right ] = \sum_{n} e^{- \epsilon_n  \tau},
 \end{equation}
where $\epsilon_n$'s are quantized eigen-energy values of  the interacting  Hamiltonian. The non-interacting integrated correlation function, $C_0 (t)$, has the same form by turning off contact interaction potential.

  \begin{figure}
\includegraphics[width=0.45\textwidth]{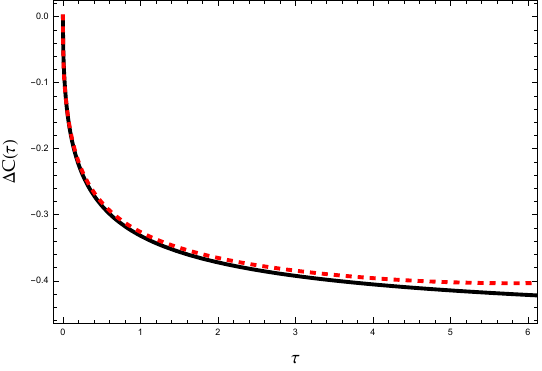}
\caption{  $\Delta C (\tau)= C(\tau) - C_0 (\tau)$  with  exact solution (solid black) given by Eq.(\ref{eq:dctexactsol})  vs. the  results (red dashed)  by analytical continuation  of   real-time evolution of $ \Delta C (t)  = Tr \left [ e^{- i \hat{H} t}  - e^{- i \hat{H}_0 t} \right ] $ by using Eq.(\ref{eq:mainresult}).   The parameters are: $V_0 = 2$,   $a_x=0.1$,  $a=1.0$ and $c=-0.1$.   The cutoff on $t$ integration is taken   $t_\mathrm{cut}=100$. The size of Hamiltonian matrix is $81\times 81$.   }\label{Fig:dctplot}
\end{figure}

For a contact interaction potential,  the analytic solution of infinite volume scattering phase shift is available,
\begin{equation}
\delta(E) = \cot^{-1} \left ( - \frac{\sqrt{2 m E} }{m V_0} \right ).
\label{eq:invPhase}
\end{equation}  
 Hence the analytic solution of Eq.(\ref{eq:ICFmainEQ}) can be obtained
 \begin{equation}
C(\tau) - C_0 (\tau)   =  \frac{1}{2} \mbox{erfc}  \left( m V_0  \sqrt{\frac{\tau}{2m}}\right) e^{ (m V_0)^2 \frac{\tau}{2m}} - \frac{1}{2} .  \label{eq:dctexactsol}
\end{equation}
 The exact solution of $\Delta C (\tau)= C(\tau) - C_0 (\tau)$     can be compared with the results by analytic continuation using Eq.(\ref{eq:mainresult}) from real-time evolution of  
 \begin{equation}
  \Delta C (t)  = C(t) - C_0 (t)= Tr \left [ e^{- i \hat{H} t} \right ]  - Tr \left [ e^{- i \hat{H}_0 t} \right ] ,
  \end{equation}
   where $\hat{H}_0 = - \frac{1}{2m}  \frac{d^2}{d x^2} $ is the non-interacting Hamiltonian. The  real-time evolution of   $ C (t)   $ and $C_0 (t)$  are solved numerically by discretizing  the coordinate $x = a_x \alpha$ where $\alpha \in \mathbb{Z}$ again.  The discretized 
Hamiltonian operator  can be written as,
\begin{equation}
\hat{H}  =  \sum_{\alpha \in \mathbb{Z}}    V(\alpha )     | \alpha \rangle \langle \alpha  |  - \frac{1}{2 m a_x^2}  \sum_{\alpha \in \mathbb{Z}}      \left ( | \alpha \rangle \langle \alpha +1 |  + | \alpha +1 \rangle \langle \alpha  |  \right )     , \label{eq:QMHmatrix}
\end{equation}
where 
\begin{equation}
V(\alpha)  =  \frac{1}{ m a_x^2}  + \dfrac{V_0}{ a_x } \delta_{\alpha, 0} .   \label{eq:HVQMdef}
\end{equation}
The demo plots of exact solutions of $\Delta C (\tau)$ given by Eq.(\ref{eq:dctexactsol})   vs. the results by    analytic continuation using Eq.(\ref{eq:mainresult}) from real-time evolution of $ C (t)   $ and $C_0 (t)$  are given in Fig.~\ref{Fig:dctplot}.  Again, the exact solutions and analytic continuation results  agree well in Fig.~\ref{Fig:dctplot} {  for small  $t$. The deviations  at large $t$ are primarily caused by both $t_{cut}$ and boundary condition effect. The agreement will be improved for larger $t_{cut}$ and larger size of Hamiltonian matrix, we refer readers to Refs.~\cite{Guo:2023ecc,Guo:2024zal,Guo:2024pvt,Guo:2025vgk,Guo:2025ngh,Guo:2025lmd}  for the complete discussion on the boundary condition effects for Eq.(\ref{eq:ICFmainEQ}). }

\section{Numerical test on IBM hardware}\label{sec:numericsIBM} 

\begin{figure}[h]
\frame{\includegraphics[width=.45\textwidth]{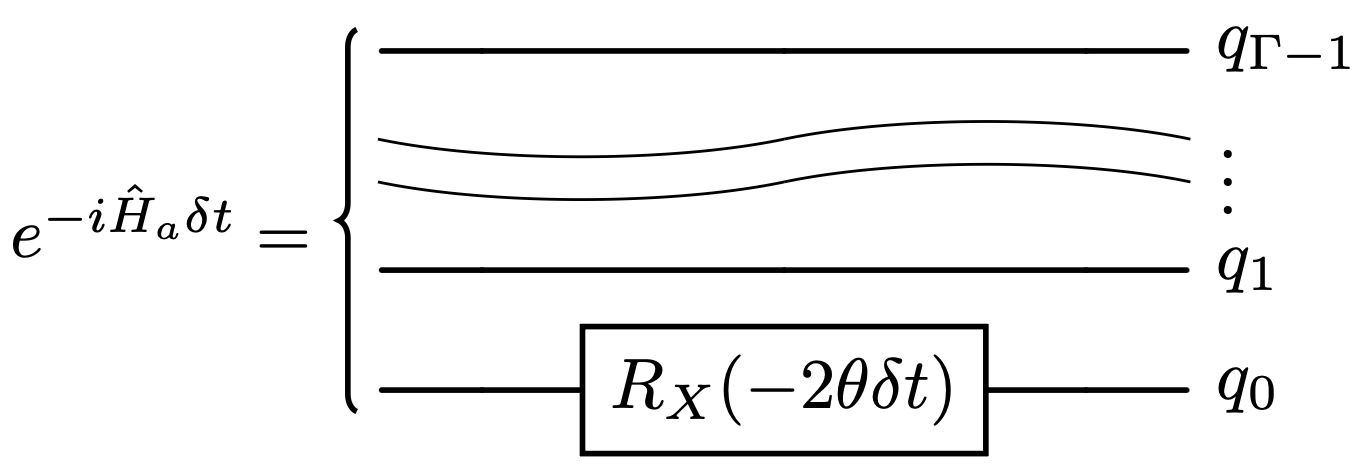}}
\caption{Quantum circuit for time evolution of $e^{- i \hat{H}_a \delta t}$.}
\label{FIGexpHa}
\end{figure}

\begin{figure}[h]
\frame{\includegraphics[width=.45\textwidth]{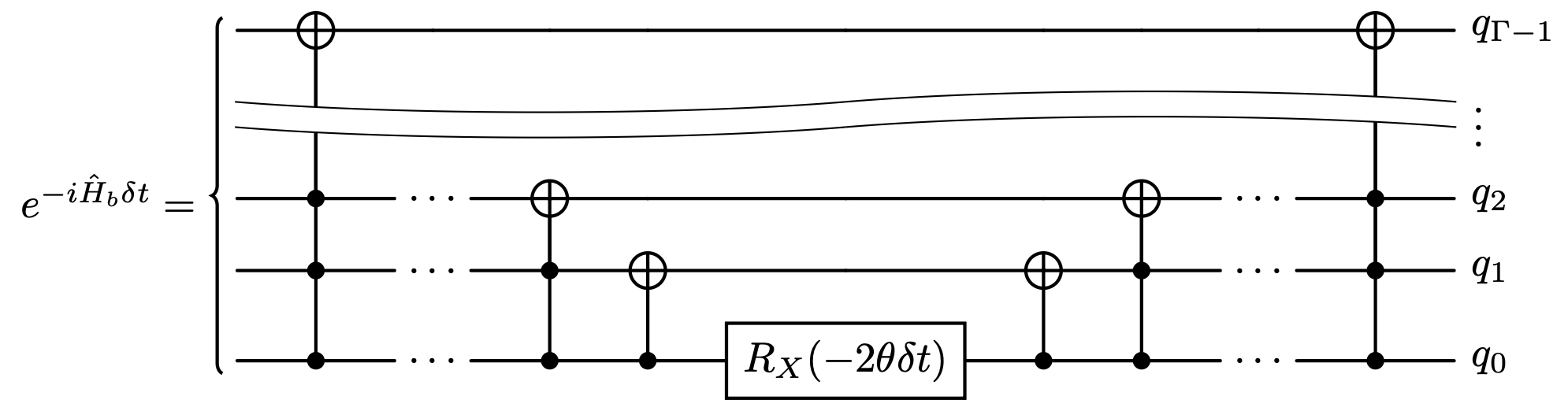}}
\caption{Quantum circuit for time evolution of $e^{- i \hat{H}_b \delta t}$.}
\label{FIGexpHb}
\end{figure}

 To implement real-time evolution given by the unitary operator $e^{- i \hat{H} t} $ on quantum hardware, let us split the Hamiltonian in Eq.(\ref{eq:FPHmatrix}) into three terms, 
\begin{align}
\hat{H} &= \hat{H}_a +\hat{H}_b + \hat{H}_v, \text{ with} \nonumber \\
\hat{H}_a  & = -  \theta \sum_{\alpha = 0}^{\frac{2^\Gamma}{2}-1}     \Bigl (\, | 2 \alpha \rangle \langle 2 \alpha +1 |  + | 2 \alpha +1 \rangle \langle 2 \alpha  |  \,\Bigr ) , \nonumber \\
\hat{H}_b  & = - \theta \sum_{\alpha = 0}^{\frac{2^\Gamma}{2}-1}     \Bigl (\, | 2 \alpha  + 1 \rangle \langle 2 \alpha +2 |  + | 2 \alpha +2 \rangle \langle 2 \alpha  + 1 |  \,\Bigr ) , \nonumber \\
\hat{H}_v  & =  \sum_{\alpha = 0}^{2^\Gamma-1}    V(\alpha )   | \alpha \rangle \langle \alpha  |  , \label{HaHbHvcoordinate}
\end{align}
where  $\Gamma$ is the number of qubits that are used to encode the dynamics of physical systems, and
\begin{equation}
\theta = \begin{cases}  \frac{  D }{a_x^2}   , &  \mbox{Fokker-Planck equations}; \\    \frac{  1 }{2 m a_x^2}   , & \mbox{scattering in QM}. \end{cases}
\end{equation}
The $ V(\alpha ) $ for  Brownian motion with a constant drift process  and   Ornstein-Uhlenbeck  Process   are defined in Eq.(\ref{eq:HVFPdef}).  For 1D scattering in quantum mechanics, the     $ V(\alpha ) $ function is shifted to the center of periodic box,
\begin{equation}
V(\alpha)  =  \frac{1}{ m a_x^2}  + \dfrac{V_0}{ 2 a_x } \left ( \delta_{\alpha, \frac{2^\Gamma }{2}-1} +\delta_{\alpha, \frac{2^\Gamma }{2}} \right ) .   \label{eq:HVQMshifteddef}
\end{equation}
The $\hat{H}_a $ and $\hat{H}_b $ terms together are exactly the same as the first two terms in a tight-binding model in Ref.~\cite{Guo:2025xpd}.  Quantum circuits of time evolution operators  $e^{- i \hat{H}_a \delta t}$ and $e^{- i \hat{H}_b \delta t}$ are given in Fig.~\ref{FIGexpHa} and Fig.~\ref{FIGexpHb}, also see details in Ref.~\cite{Guo:2025xpd}, where the periodic boundary condition has been considered: $| 0\rangle = | 2^\Gamma \rangle$.   The  $\hat{H}_v $ is a diagonal matrix, the $e^{- i \hat{H}_v \delta  t}$ is thus   a diagonal unitary matrix and can be coded easily by converting an unitary matrix into a quantum circuit using operator class in IBM Qiskit \cite{Javadi-Abhari:2024kbf}. 
The real time evolution of the total Hamiltonian matrix can be  computed via the Trotterization approximation \cite{Trotter1959,Hatano:2005gh}, e.g. the lowest order of  Trotterization  is given by
\begin{equation}
e^{- i \hat{H}   t  } \stackrel{ \delta t \rightarrow 0}{\approx }  \left [  e^{- i  \hat{H}_{v} \delta t}  e^{- i  \hat{H}_{b} \delta t} e^{- i  \hat{H}_{a} \delta t}   \right ]^{N_t}    ,
\end{equation}
 where $t = \delta t N_t$.

\begin{figure}[h]
\frame{\includegraphics[width=.45\textwidth]{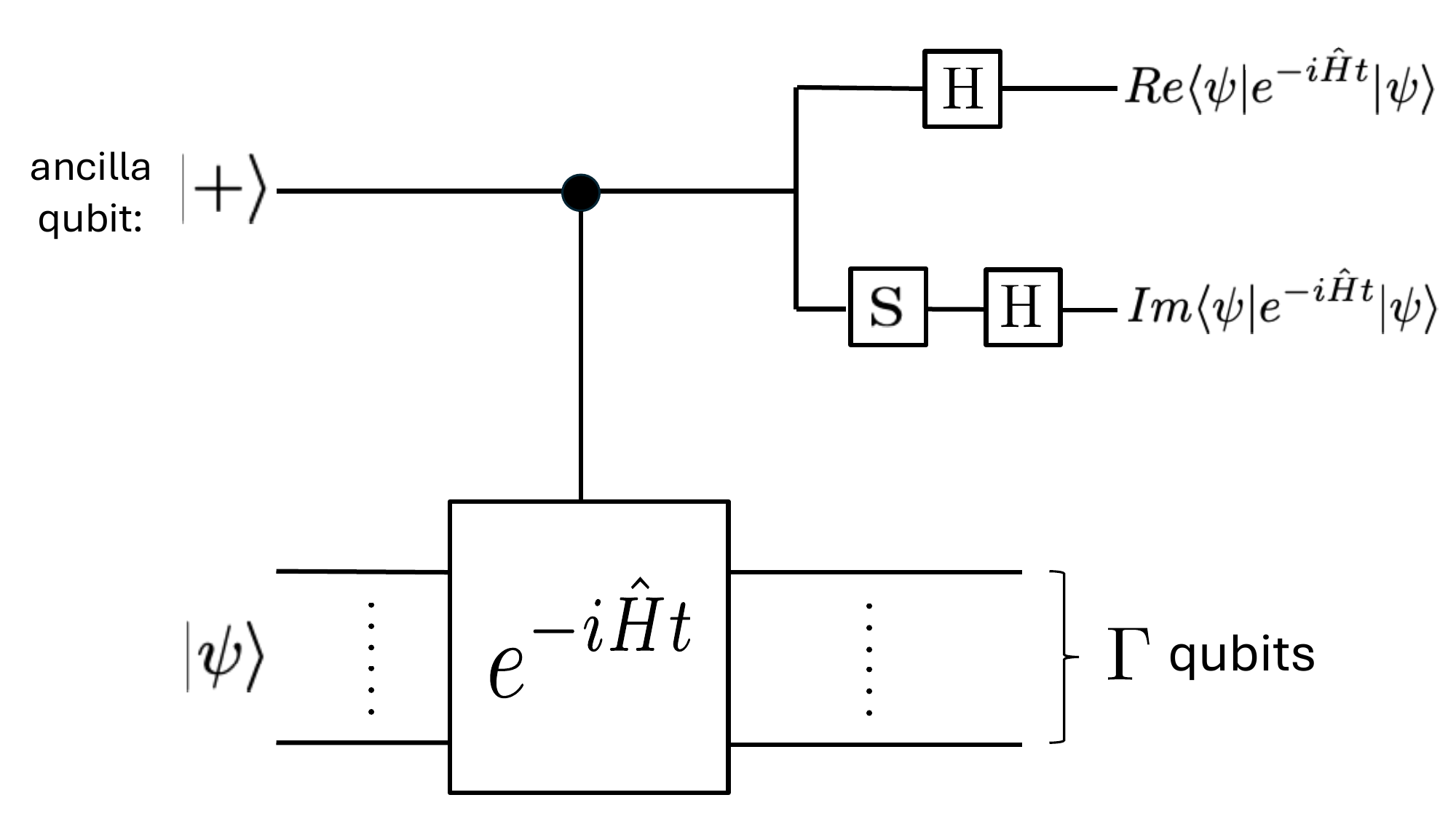}}
\caption{Quantum circuit for computing the integrated correlation function $  \langle \psi | e^{- i \hat{H} t} | \psi \rangle$. }
\label{FIGctQC}
\end{figure}

\paragraph{Computing expectation value of $\langle \psi | e^{- i \hat{H} t}  | \psi  \rangle$ on quantum computers: }  For an arbitrary quantum state  $| \psi \rangle$,  the expectation value of real-time evolution operator,  $\langle \psi | e^{- i \hat{H} t} | \psi \rangle$, can be computed  by   Hadamard test method \cite{PhysRevA.75.012328,PhysRevD.102.094505}, also see Fig.~\ref{FIGctQC}.   The real part of $  \langle \psi | e^{- i \hat{H} t} | \psi \rangle$ is given  by $P(0) - P(1)$ in  upper branch of ancillary qubit measurements in Fig.~\ref{FIGctQC}, where $P(0)$ and $P(1)$ are the probabilities of ancillary qubit at state $|0 \rangle $ and $|1\rangle $ respectively.  The imaginary part of $\langle \psi | e^{- i \hat{H} t} | \psi  \rangle$ is given  by $P(1) - P(0)$ in lower branch of ancillary qubit measurements in Fig.~\ref{FIGctQC}. The real-time evolution of integrated correlation functions 
\begin{equation}
C(t) = \sum_{\alpha = 0}^{2^\Gamma -1} \langle \alpha | e^{- i \hat{H} t} | \alpha \rangle  \ \ \mbox{and}  \ \  C_0(t) = \sum_{\alpha = 0}^{2^\Gamma -1} \langle \alpha | e^{- i \hat{H}_0 t} | \alpha \rangle
\end{equation}
  can be computed on quantum computers using quantum circuit shown in Fig.~\ref{FIGctQC}, see e.g. Refs.~\cite{Guo:2026qkx,Guo:2026nuc}.

\paragraph{Computing matrix element of $\langle \alpha | e^{- i \hat{H} t}  | e^{\frac{\Phi}{2}}W(0) \rangle$ on quantum computers: } The matrix element of  $\langle \alpha | e^{- i \hat{H} t}  | e^{\frac{\Phi}{2}}W(0) \rangle$  amplitude can be constructed by computation of four separate expectation values of real-time evolution operator through $|  \psi_\pm^{(X)}  \rangle$ and $|  \psi_\pm^{(Y)}  \rangle$ four states, where
 \begin{align}
|  \psi_\pm^{(X)}  \rangle  &=  \frac{  |  \alpha  \rangle \pm | e^{\frac{\Phi}{2}}W(0) \rangle  }{\sqrt{2}} , \nonumber \\
    | \psi_\pm^{(Y)} \rangle  & =  \frac{  |  \alpha  \rangle \pm i | e^{\frac{\Phi}{2}}W(0) \rangle  }{\sqrt{2}}   ,
\end{align}
 by using relation  \cite{Kirby:2025iiw},
\begin{align}
& \langle \alpha | e^{- i \hat{H} t}  | e^{\frac{\Phi}{2}}W(0) \rangle \nonumber \\
& =\frac{ 1  }{2} \left (  \langle  \psi^{(X)}_+    | e^{- i \hat{H} t}  |  \psi^{(X)}_+   \rangle  - \langle  \psi_-^{(X)}    | e^{- i \hat{H} t}  |  \psi_-^{(X)}   \rangle  \right ) \nonumber \\
& + \frac{ i  }{2} \left ( \langle  \psi_+^{(Y)}    | e^{- i \hat{H} t}  |  \psi_+^{(Y)}   \rangle  - \langle  \psi_-^{(Y)}    | e^{- i \hat{H} t}  |  \psi_-^{(Y)}   \rangle  \right )    .  \label{eq:offdiagonalamp}
\end{align}
In the applications presented here, since $\hat{H}$ is completely real,  the matrix element of  $\langle \alpha | e^{- i \hat{H} t}  | e^{\frac{\Phi}{2}}W(0) \rangle$   is transpose symmetric:
\begin{equation}
\langle \alpha | e^{- i \hat{H} t}  | e^{\frac{\Phi}{2}}W(0) \rangle = \langle e^{\frac{\Phi}{2}}W(0)  | e^{- i \hat{H} t}  | \alpha \rangle ,
\end{equation}
hence the last two terms in Eq.(\ref{eq:offdiagonalamp}) cancel out. The construction of   the matrix element of  $\langle \alpha | e^{- i \hat{H} t}  | e^{\frac{\Phi}{2}}W(0) \rangle$  amplitude  only require two separate expectation values of real-time evolution operator through $|  \psi_\pm^{(X)}  \rangle$, 
\begin{align}
& \langle \alpha | e^{- i \hat{H} t}  | e^{\frac{\Phi}{2}}W(0) \rangle \nonumber \\
& =\frac{ 1  }{2} \left (  \langle  \psi^{(X)}_+    | e^{- i \hat{H} t}  |  \psi^{(X)}_+   \rangle  - \langle  \psi_-^{(X)}    | e^{- i \hat{H} t}  |  \psi_-^{(X)}   \rangle  \right )    .  \label{eq:offdiagonalamprealH}
\end{align}

\subsection{Brownian motion with a constant drift on IBM quantum hardware}

For the numerical test on IBM quantum hardware, we chose a normalized initial state $| e^{ \frac{\Phi }{2} }W(  0) \rangle$ for Brownian motion with a constant drift process,
\begin{equation}
\langle \alpha | e^{ \frac{\Phi }{2} }W(  0) \rangle   =\frac{1  }{   \sqrt{ N_{e^{\frac{\Phi}{2}}W} } } \begin{cases}  e^{ - \frac{\gamma a_x \alpha}{2 D} }   , & \alpha  \in [ \frac{2^\Gamma}{2}-1 ,  \frac{2^\Gamma}{2} ] \\ 0, & \text{otherwise}  \end{cases}    . \label{eq:expPhiW0IBM}
\end{equation}
where the normalization factor is defined by
\begin{equation}
N_{e^{\frac{\Phi}{2}}W}  = \sum_{ \alpha  \in [ \frac{2^\Gamma}{2}-1 ,  \frac{2^\Gamma}{2} ]   } e^{ - \frac{\gamma a_x \alpha}{ D} }    .
\end{equation}
The center of rectangle shape of state $| W(0)\rangle$ itself is placed in the middle of a periodic box with range in $x = a_x \alpha \in [0, L]$, and the width of rectangle shape of state $| W(0)\rangle$ itself is $b= 2 a_x$. The real-time evolution of matrix element 
\begin{align}
& \langle \alpha | e^{- i \hat{H} t}  | e^{\frac{\Phi}{2}}W(0) \rangle  \nonumber \\
& \approx  e^{- i  \left ( \frac{2 D}{a_x^2}  + \frac{  \gamma^2  }{4D}   \right )  t}  \langle \alpha | \left [  e^{- i \hat{H}_b \delta t}  e^{- i \hat{H}_a \delta t}  \right ]^{N_t} | e^{\frac{\Phi}{2}}W(0) \rangle
\end{align}
can thus be evaluated on quantum hardware by using Eq.(\ref{eq:offdiagonalamprealH}), where the quantum circuits of  $e^{- i \hat{H}_a \delta t}$ and $e^{- i \hat{H}_b \delta t}$ are defined  in Fig.~\ref{FIGexpHa} and Fig.~\ref{FIGexpHb} respectively.

\begin{figure}[h]
\includegraphics[width=.45\textwidth]{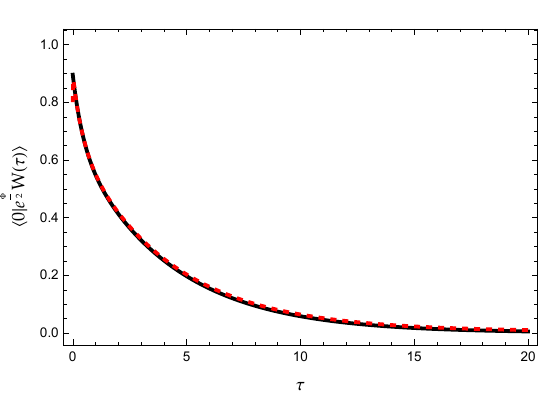}
\includegraphics[width=.45\textwidth]{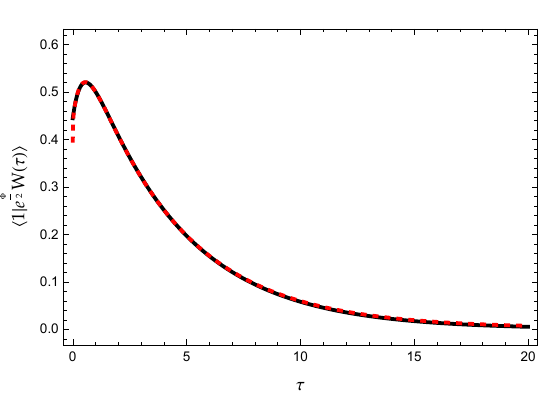}
\caption{Comparison between exact solutions (solid black) of $  \langle \alpha | e^{-  \hat{H} \tau} | e^{\frac{\Phi}{2}} W(0) \rangle$ vs. the results (dashed red) by analytic continuation by using Eq.(\ref{eq:BMnumericresult}) with 1-qubit for Brownian motion with a constant drift.  The  matrix elements $  \langle \alpha | e^{-  i \hat{H} t} | e^{\frac{\Phi}{2}} W(0) \rangle$ are computed on IBM quantum hardware with 1-qubit, the results   for $\alpha =0,1$ are plotted in upper and lower panels respectively.     The model parameters are: $\gamma = 1.6$, $D = 2.0$, $L=4.0$, $a_x = \frac{L}{2}=2.0$,  $a=1.0$ and $c=-0.001$.  The cutoff of $t$ integral in Eq.(\ref{eq:I12BMnumerics}) is $t_\mathrm{cut}=200$, and $400$ weights and nodes of Gaussian quadrature are generated.}
\label{FIGITEIBM1qubitBMplots}
\end{figure}

\begin{figure}[h]
\includegraphics[width=.45\textwidth]{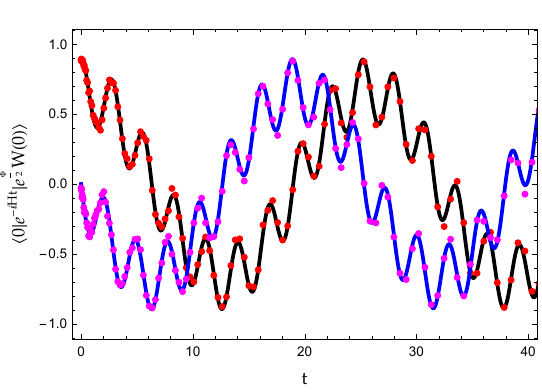}
\includegraphics[width=.45\textwidth]{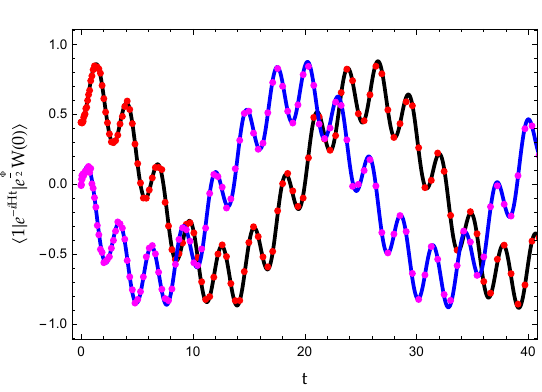}
\caption{Comparison between exact solutions (solid black for real part and solid blue for imaginary part) of $  \langle \alpha | e^{- i \hat{H} t} | e^{\frac{\Phi}{2}} W(0) \rangle$ vs. the results from IBM quantum hardware (red dots for real part and purple dots for imaginary parts)  with 1-qubit for Brownian motion with a constant drift.   The results   for $\alpha =0,1$ are plotted in upper and lower panels respectively.     The model parameters are same as  parameters in Fig.~\ref{FIGITEIBM1qubitBMplots}. The number of shots are 4000 for quantum measurements. }
\label{FIGRTEIBM1qubitBMplots}
\end{figure}

The imaginary-time evolution of matrix element  $\langle \alpha | e^{-  \hat{H} \tau }  | e^{\frac{\Phi}{2}}W(0) \rangle$ thus can be obtained by using analytic continuation relation given in Eq.(\ref{eq:mainresult}) from  results of real-time evolution of matrix element $\langle \alpha | e^{- i \hat{H} t}  | e^{\frac{\Phi}{2}}W(0) \rangle $ on IBM quantum hardware: 
\begin{align}
&  \langle \alpha | e^{-  \hat{H} \tau }  | e^{\frac{\Phi}{2}}W(0) \rangle \nonumber \\
 & =  - \frac{1}{\pi }    \int_{0 }^{ a}   \mathrm{d} \epsilon\   e^{- c \tau}   \ \mathrm{Im}  \left [ e^{- i \epsilon  \tau}  \langle \alpha |   \hat{I}_1 (\epsilon) | e^{\frac{\Phi}{2}}W(0) \rangle   \right ]  \nonumber \\
 & +  \frac{1}{\pi }   \int_{c }^{  \infty }   \mathrm{d} \epsilon\   e^{- \epsilon   \tau} \ \mathrm{Re} \left [  e^{-    i a \tau} \langle \alpha | \hat{I}_2 (\epsilon)   | e^{\frac{\Phi}{2}}W(0) \rangle  \right ] . \label{eq:BMnumericresult}
\end{align}
where we have utilised the fact that $\hat{H}$ is real to simplify the expressions. The elements of $ \hat{I}_{1,2} (\epsilon)$ functions in Eq.(\ref{eq:BMnumericresult}) are evaluated by Gaussian quadrature approximation
\begin{align}
 \langle \alpha |   \hat{I}_1 (\epsilon) | e^{\frac{\Phi}{2}}W(0) \rangle   & \approx  \sum_{i}  w_i e^{i (c+i \epsilon)  t_i} \langle \alpha |   e^{- i \hat{H} t_i } | e^{\frac{\Phi}{2}}W(0) \rangle   , \nonumber \\
\langle \alpha | \hat{I}_2 (\epsilon)   | e^{\frac{\Phi}{2}}W(0) \rangle& \approx  \sum_{i}  w_i  e^{i (\epsilon + i a) t_i} \langle \alpha |  e^{- i \hat{H} t_i }  | e^{\frac{\Phi}{2}}W(0) \rangle    , \label{eq:I12BMnumerics}
\end{align}
where $w_i$ and $t_i$ are the weights and nodes of Gaussian quadrature method. In Fig.~\ref{FIGITEIBM1qubitBMplots}, we show the 1-qubit exact  results of matrix element  $\langle \alpha | e^{-  \hat{H} \tau }  | e^{\frac{\Phi}{2}}W(0) \rangle$ vs. the results by analytic continuation  by using Eq.(\ref{eq:BMnumericresult}).  The  matrix elements$  \langle \alpha | e^{-  i \hat{H} t} | e^{\frac{\Phi}{2}} W(0) \rangle$ are computed on IBM quantum hardware with 1-qubit for  400 Gaussian quadrature nodes $t_i \in [0,200]$,    demo plot of exact  results of matrix element  $\langle \alpha | e^{-  i  \hat{H} t }  | e^{\frac{\Phi}{2}}W(0) \rangle$ vs. the results from IBM quantum hardware are shown in  Fig.~\ref{FIGRTEIBM1qubitBMplots}. As we can see in Fig.~\ref{FIGITEIBM1qubitBMplots}, the exact result of matrix element  $\langle \alpha | e^{-  \hat{H} \tau }  | e^{\frac{\Phi}{2}}W(0) \rangle$ agree reasonably well with the results by analytic continuation even with the cutoff at $200$ on the $t$ integral in Eq.(\ref{eq:BMnumericresult}), the agreement can be further improved if the cutoff is pushed further.

\begin{figure}[h]
\includegraphics[width=.45\textwidth]{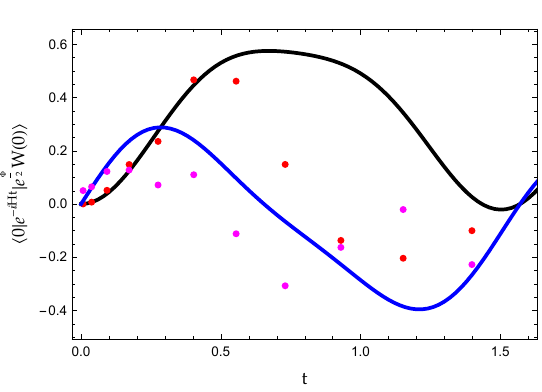}
\caption{Comparison between exact solutions (solid black for real part and solid blue for imaginary part) of $  \langle \alpha | e^{- i \hat{H} t} | e^{\frac{\Phi}{2}} W(0) \rangle$ with $\alpha =0$  vs. the results from IBM quantum hardware (red dots for real part and purple dots for imaginary parts)  with 2-qubit for Brownian motion with a constant drift.       The model parameters are same as  parameters in Fig.~\ref{FIGITEIBM1qubitBMplots}. Shot-noise error (not shown) is not a dominant source of error. }
\label{FIGRTEIBM2qubitBMplots}
\end{figure}

Due to the high fidelity ($\sim 99.976\%$) of IBM single-qubit gate on  Heron r3 QPUs, the 1-qubit results shown in Fig.~\ref{FIGITEIBM1qubitBMplots} has been encouraging,  the trotter approximation can be even pushed up to 2000 time steps. Unfortunately, without noise-mitigation strategies, the real-time evolution results on the 2-qubit system is ill-constrained after very short times, as shown in Fig.~\ref{FIGRTEIBM2qubitBMplots}. The primary causes of failure are  due to two-qubit gate operation errors and thermal relaxation errors, see detailed discussions in Ref.~\cite{Guo:2026qkx}.

\subsection{Imaginary-time evolution of correlation functions in 1D quantum mechanical scattering on IBM quantum hardware}

\begin{figure}[h]
\includegraphics[width=.45\textwidth]{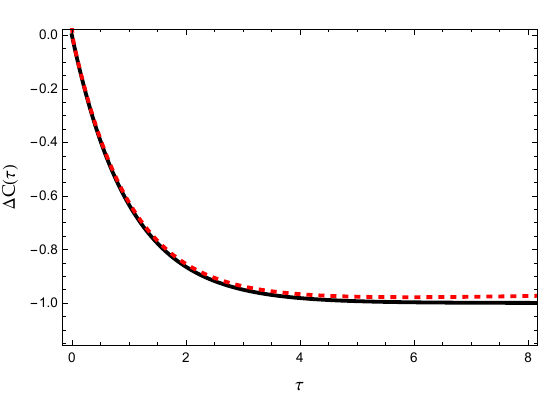}
\caption{Comparison between exact solutions (solid black) of $ \Delta C(\tau) = \mathrm{Tr}\left [ e^{- \hat{H} \tau} - e^{- \hat{H}_0 \tau} \right ] $ vs. the results (dashed red) by analytic continuation by using Eq.(\ref{eq:CTnumericresult}) with 1-qubit.  The   $ \Delta C(t) = \mathrm{Tr}\left [ e^{- i \hat{H} t} - e^{- i  \hat{H}_0 t} \right ] $  are computed on IBM quantum hardware with 1-qubit.     The model parameters are:  $L=4.0$, $a_x = \frac{L}{2}=2.0$, $V_0 =2$,  $a=1.0$ and $c=-0.2$.  The cutoff of $t$ integral in Eq.(\ref{eq:I12CTnumerics}) is $t_\mathrm{cut}=100$ and $200$ weights and nodes of Gaussian quadrature are used.}
\label{FIGITEIBM1qubitdCtauplots}
\end{figure}

\begin{figure}[h]
\includegraphics[width=.45\textwidth]{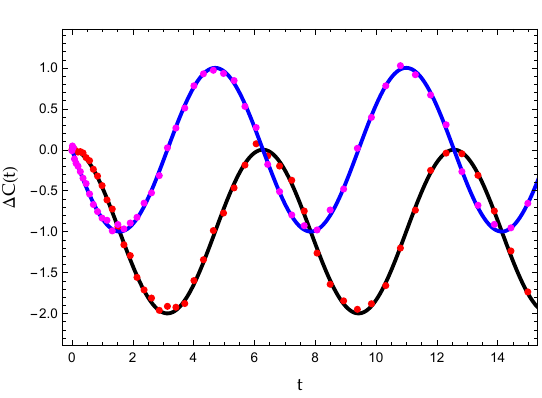}
\caption{Comparison between exact solutions (solid black for real part and solid blue for imaginary part) of  $ \Delta C(t) = Tr\left [ e^{- i \hat{H} t} - e^{- i  \hat{H}_0 t} \right ] $   vs. the results from IBM quantum hardware (red dots for real part and purple dots for imaginary parts)  with 1-qubit for quantum mechanical scattering. The model parameters are same as  parameters in Fig.~\ref{FIGITEIBM1qubitdCtauplots}. The number of shots are 2000 for quantum measurements, such that shot-noise is not a dominant source of error. }
\label{FIGRTEIBM1qubitdCtplots}
\end{figure}

Next, we perform the second numerical test on IBM quantum hardware with the imaginary-time evolution of integrated correlation functions in 1D quantum mechanical scattering that is described in Sec.~\ref{sec:1Dctquantum}. The real-time evolution of $C(t) =\mathrm{Tr} \left [e^{- i \hat{H} t} \right ]$ and $C_0(t) =\mathrm{Tr} \left [e^{- i \hat{H}_0 t} \right ]$ can be computed rather straight-forwardly using  quantum circuit   in Fig.~\ref{FIGctQC}, more details are presented in Refs.~\cite{Guo:2026qkx,Guo:2026nuc}.  The imaginary-time evolution of  $C(\tau) =\mathrm{Tr} \left [e^{-  \hat{H} \tau} \right ]$ and  $C_0(\tau) =\mathrm{Tr} \left [e^{-  \hat{H}_0 \tau} \right ]$ thus can be obtained by analytic continuation (and utilising the fact that $\hat{H}$ is a real matrix):
\begin{align}
 C(\tau )  
 & =  - \frac{1}{\pi }    \int_{0 }^{ a}   \mathrm{d} \epsilon\   e^{- c \tau}   \ \mathrm{Im}  \left [ e^{- i \epsilon  \tau}   \hat{I}^{(C)}_1 (\epsilon)  \right ]  \nonumber \\
 & +  \frac{1}{\pi }   \int_{c }^{  \infty }   \mathrm{d} \epsilon\   e^{- \epsilon   \tau} \ \mathrm{Re} \left [  e^{-    i a \tau}   \hat{I}^{(C)}_2 (\epsilon)   \right ] , \label{eq:CTnumericresult}
\end{align}
where
\begin{align}
   I^{(C)}_1 (\epsilon)   & \approx  \sum_{i}  w_i e^{i (c+i \epsilon)  t_i}     C(t_i)  , \nonumber \\
 I^{(C)}_2 (\epsilon)  & \approx  \sum_{i}  w_i  e^{i (\epsilon + i a) t_i}     C(t_i)   . \label{eq:I12CTnumerics}
\end{align}
 In Fig.~\ref{FIGITEIBM1qubitdCtauplots}, we show the 1-qubit exact  results of  $ \Delta C(\tau) = \mathrm{Tr}\left [ e^{- \hat{H} \tau} - e^{- \hat{H}_0 \tau} \right ] $  vs. the results by analytic continuation  by using Eq.(\ref{eq:CTnumericresult}).  The   $ \Delta C(t) = \mathrm{Tr}\left [ e^{- i \hat{H} t} - e^{- i  \hat{H}_0 t} \right ] $   are computed on IBM quantum hardware with 1-qubit for  200 Gaussian quadrature nodes $t_i \in [0,100]$,    demo plot of exact  results of  $ \Delta C(t)  $   vs. the results from IBM quantum hardware are shown in  Fig.~\ref{FIGRTEIBM1qubitdCtplots}.    Again with   $t$-integration cutoff to $100$, the agreement between exact result of $\Delta C(\tau) $  vs.  the result by analytic continuation   are not bad for 1-qubit results. The results can be improved by pushing the $t$-integration cutoff to higher value.

\section{Summary and discussion}\label{sec:summary}

In this work, a method to obtain imaginary-time evolution directly by analytic continuation of real-time correlation functions was explored. 
Compared to other approaches - this approach is simple to implement, requiring only a lower bound $c$ on the ground state energy of the Hamiltonian, and the ability to efficiently perform real-time evolution of the system. 

The  feasibility  of analytic continuation method  are demonstrated  with three  exactly solvable   1D examples: (1) Brownian motion with a constant drift; (2) Ornstein-Uhlenbeck Process; and (3) scattering  in quantum mechanics. The numerical tests are carried out on IBM quantum hardware. The 1-qubit results are positive and encouraging due to high fidelity of single-qubit gate of IBM Heron r3 QPUs, however 2-qubit results fails after only a few trotter steps    due to two-qubit gate operation errors and thermal relaxation errors.
Without further error mitigation strategies, it will be difficult to proceed to systematically controlled calculations on larger quantum systems. 

Although the examples considered in this work have focused on 1D Schr\"odinger-equation-like systems, the analytic continuation technique should be applicable to arbitrary Hamiltonians. 
The specific case of non-relativistic field theory dynamics is discussed in more detail in Appendix~\ref{sec:appendix-manybody-application}.

\acknowledgments
 
This research is supported by the U.S. National Science Foundation under grant PHY-2418937 (P.G.) and PHY-2531653 (P.G.), and by the U.S. Department of Energy under grant   DE-AC02-06CH11357 (J.L., Y.Z.) and DOE Early Career Award through Contract No.~DE-SCL0000017 (J.L., Y.Z.).

\appendix

\section{Potential applications in many-body theory}\label{sec:appendix-manybody-application}

Although our discussion in the main text is only limited to the quantum mechanical like applications, the analytic continuation technique should be applicable to non-relativistic field theory dynamics in general, such as many-body theory.   Let's use a    single fermion's correlation  function in many-body theory  as an example.     The  single fermion's correlation function is defined by~\cite{fetter2003quantum},
\begin{equation}
   C_{\alpha, \beta} (x , t ; x' , 0)  =   \langle \Psi_0 |  T \left [ \hat{\psi}_{H \alpha} (x, t)   \hat{\psi}^\dag_{H \beta} (x', 0) \right ]| \Psi_0 \rangle  ,  
\end{equation}
where  $\hat{\psi}_{H \alpha} (x, t)  =e^{ i \hat{H} t}  \hat{\psi}_{ \alpha} (x)  e^{- i \hat{H} t}$ is    single fermion  Heisenberg field operator. Spin polarization of the fermion operator is labeled by $(\alpha, \beta)$ indices, and $ | \Psi_0 \rangle $ refers to the ground state of a many-body system with ground state energy of $E_0$. The time ordered single fermion's correlation function can be split up into two pieces,
\begin{align}
& C_{\alpha, \beta} (x , t ; x' , 0)  \nonumber \\
& = \theta(t) \langle \Psi_0 | \hat{\psi}_\alpha (x) e^{- i  (\hat{H}- E_0) t }  \hat{\psi}^\dag_\beta (x')| \Psi_0 \rangle   \nonumber \\
&  - \theta(-t) \langle \Psi_0 |    \hat{\psi}^\dag_\beta (x') e^{ i ( \hat{H} -E_0) t }  \hat{\psi}_\alpha (x) | \Psi_0 \rangle   ,
\end{align}
where one is related to the  particles    propagating time-forward and another is associated with holes    propagating time-backward: 
\begin{align}
& C_{\alpha, \beta} (x , t ; x' , 0)  \nonumber \\
& = \theta (t)  \sum_n   \langle \Psi_0 | \hat{\psi}_\alpha (x)  | \Psi_n \rangle \langle \Psi_n  | \hat{\psi}^\dag_\beta (x')| \Psi_0 \rangle  e^{- i   (E_n - E_0) t }  \nonumber \\
&- \theta (-t)  \sum_n   \langle \Psi_0 |    \hat{\psi}^\dag_\beta (x') | \Psi_n \rangle \langle \Psi_n  |   \hat{\psi}_\alpha (x) | \Psi_0 \rangle  e^{ i   (E_n - E_0) t } ,
\end{align}
where $ | \Psi_n \rangle  $ and $E_n$ are the eigen-solutions of many-body system Hamiltonian.  The full Green's function is typically defined by
\begin{equation}
  G_{\alpha, \beta} (x , x'  ; E)  = - i \int_{-\infty}^\infty  \mathrm{d} t\   e^{i E t}  C_{\alpha, \beta} (x , t ; x' , 0)   , 
\end{equation}
so that many-body Green's function has poles located at both along positive and negative real axis on complex energy plane that are related to particles    propagating time-forward and   holes    propagating time-backward respectively,
\begin{align}
 & G_{\alpha, \beta} (x , x'  ; E)   \nonumber \\
 & =     \sum_n   \frac{   \langle \Psi_0 | \hat{\psi}_\alpha (x)  | \Psi_n \rangle \langle \Psi_n  |  \hat{\psi}^\dag_\beta (x')| \Psi_0 \rangle   }{E -   (E_n - E_0) }   \nonumber \\
 & +  \sum_n    \frac{  \langle \Psi_0 |    \hat{\psi}^\dag_\beta (x')   | \Psi_n \rangle \langle \Psi_n  | \hat{\psi}_\alpha (x) | \Psi_0 \rangle   }{E + ( E_n  -E_0) }   . \label{eq:fullGreenManyBody}
\end{align}

The analytic continuation technique that is described in  Sec.~\ref{sec:analycontintro}  is thus still applicable in many-body theory, however we now need to be mindful on the choice of contour in Fig.~\ref{Fig:contourintegral} to make sure the particle-poles and hole-poles are distributed on the two sides of contour $C$ in $\int_C \mathrm{d} E e^{- E \tau} [\cdots ]$ integration. For instance, the imaginary-time evolution of retarded single fermion correlation $(\tau >0)$,
 \begin{equation}
  C^{(R)}_{\alpha, \beta} (x ,  \tau   ; x' , 0)   =   \langle \Psi_0 | \hat{\psi}_\alpha (x)     e^{- ( \hat{H} - E_0 ) \tau } \hat{\psi}^\dag_\beta (x')| \Psi_0 \rangle     ,
\end{equation}
can be computed from  full many-body  Green's function in Eq.(\ref{eq:fullGreenManyBody}) by
   \begin{align}
  C^{(R)}_{\alpha, \beta} (x ,  \tau   ; x' , 0)   =   \frac{i}{2\pi } \int_{C} \mathrm{d} E   e^{- E \tau}     G_{\alpha, \beta} (x , x'  ; E)   .
\end{align}
Since  all the particle poles are on one side and all the hole poles are on the other side of contour $C$, the deformed contour $C$ in Fig.~\ref{Fig:contourintegral} will only pick up particle poles,
\begin{align}
  \frac{i}{2\pi } \int_C \mathrm{d} E  \frac{e^{- E \tau} }{E-  (E_n  - E_0)}    &=   e^{-   (E_n - E_0 ) \tau }    , \nonumber \\
  \frac{i}{2\pi } \int_C \mathrm{d} E    \frac{e^{- E \tau} }{E+  (E_n - E_0)}    &=  0   ,
\end{align}
hence we obtain, 
 \begin{align}
&  C^{(R)}_{\alpha, \beta} (x ,  \tau   ; x' , 0)   \nonumber \\
& =    \sum_n   \langle \Psi_0 | \hat{\psi}_\alpha (x)  | \Psi_n \rangle \langle \Psi_n  | \hat{\psi}^\dag_\beta (x')| \Psi_0 \rangle  e^{-    (E_n - E_0) \tau } .
\end{align}
Equivalently, instead of using full many-body Green's function in Eq.(\ref{eq:fullGreenManyBody}), we can also  project out the time propagating forward particle's Green's function  only by
\begin{align}
 &  G^{(R)}_{\alpha, \beta} (x , x'  ; E)   = - i \int_{0}^\infty  \mathrm{d} t  e^{i E t}  C_{\alpha, \beta} (x , t ; x' , 0)    \nonumber \\
 &  =     \sum_n    \frac{   \langle \Psi_0 | \hat{\psi}_\alpha (x)  | \Psi_n \rangle \langle \Psi_n  |  \hat{\psi}^\dag_\beta (x')| \Psi_0 \rangle   }{E -   (E_n - E_0) }   .
\end{align}  
 Hence contour integration $ \frac{i}{2\pi } \int_C \mathrm{d} E   e^{- E \tau}     G^{(R)}_{\alpha, \beta} (x , x'  ; E)  $ thus again yield the  imaginary-time evolution of retarded single fermion correlation.

\section{Fokker-Planck equations in 1D}\label{sec:appendix-FPequation}

 In this section, we  briefly summarize some basic results of Fokker-Planck equations in one spatial dimension or one variable that are needed in the discussion of our main texts, we refer readers to  references such as \cite{risken1989fpe,vanKampen1992} for the more technical details. In this work,  the focus will be given to  the one-variable Fokker-Planck equation that  the probability density function 
 \begin{equation}
 W(x, \tau) = \langle x | W(\tau) \rangle
 \end{equation}
  for a diffusion process is described by  a imaginary-time-Schr\"odinger-equation-like differential equation,
  \begin{equation}
 \frac{\partial}{\partial \tau } | W( \tau)  \rangle  = - \hat{L}  | W( \tau)  \rangle , \label{eq:FPLequation}
 \end{equation}
 where  $\hat{L}  $ is a non-Hermitian operator of type:
  \begin{equation}
\hat{L}  =      \frac{\partial}{\partial x}   \gamma (x)   -  D \frac{\partial^2}{\partial x^2}  . \label{eq:FPLdef}
 \end{equation}
 The $\gamma(x)$ and $D >0 $ are referred as drift and diffusion coefficients  respectively in a diffusion process, see e.g. \cite{risken1989fpe,vanKampen1992}.  The probability density function $W(x, \tau)$ must be normalized to one: 
 \begin{equation}
 \int \mathrm{d} x \ W(x,\tau) =1
 \end{equation}
  at any time $\tau$ because of probability conservation.

  The time evolution state $| W(\tau) \rangle$ is thus given by 
   \begin{equation}
  | W( \tau)  \rangle  = e^{-  \hat{L}  \tau } | W( 0)  \rangle .\label{eq:FPLtimeevolution}
 \end{equation}
 We remark that we will adopt $\tau$ to represent the physical time in a diffusion process in Fokker-Planck equation throughout this work, since Eq.(\ref{eq:FPLequation}) resemble the Schr\"odinger equations in imaginary time mathematically and can be solved through the analytic continuation from real-time evolution on quantum computers. The transition probability distribution of Fokker-Planck equation is defined by
    \begin{equation}
 P(x, \tau; x', 0)  =  \langle x | e^{-  \hat{L}  \tau } | x' \rangle  \stackrel{\tau \rightarrow 0}{\longrightarrow}  \delta(x-x') , \label{eq:FPLtransitionprob}
 \end{equation}
 hence
   \begin{equation}
   W( x, \tau)     =   \int d x' P(x, \tau  ; x' , 0)   W(x',0)   . \label{eq:Wtauevolution}
 \end{equation}

One-variable Fokker-Planck equation with a non-Hermitian operator $\hat{L}$ can be   mapped to a Schr\"odinger equation with a Hermitian Hamiltonian operator by a transformation, see e.g. Ref.~\cite{risken1989fpe},  
\begin{equation}
  \hat{L}    = e^{ - \frac{\Phi (x)}{2}}  \hat{H}  e^{ \frac{\Phi (x)}{2}}   , 
\end{equation}
where
\begin{equation}
\Phi (x) = - \int_0^x \frac{ \gamma (x')  }{D}d x' .
\end{equation}
The Eq.(\ref{eq:FPLequation}) is thus turned into an imaginary-time Schr\"odinger equation:
  \begin{equation}
 \frac{\partial}{\partial \tau } |  e^{\frac{\Phi}{2}} W( \tau)  \rangle  = - \hat{H}  |  e^{\frac{\Phi}{2}} W( \tau)  \rangle , \label{eq:FPHequation}
 \end{equation}
 where
 \begin{equation}
  \hat{H}    = -    D \frac{\partial^2}{\partial x^2}   +    \frac{\gamma^2(x)}{4D}   +  \frac{1}{2}   \frac{\partial  \gamma(x) }{\partial x} = \hat{H}^\dag  .
\end{equation}
 Hence time evolution is now given by a Hermitian operator $\hat{H}$,
    \begin{equation}
  | e^{\frac{\Phi}{2}}  W( \tau)  \rangle  = e^{-  \hat{H}  \tau } | e^{\frac{\Phi}{2}} W( 0)  \rangle .\label{eq:FPHtimeevolution}
 \end{equation}
 In terms of Hermitian operator $\hat{H}$, the   transition probability distribution   is  now given by
    \begin{equation}
 P(x, \tau; x', 0)  = e^{ - \frac{\Phi (x)}{2}}   \langle x | e^{-  \hat{H}  \tau } | x' \rangle  e^{ \frac{\Phi (x')}{2}}   . \label{eq:FPHtransitionprob}
 \end{equation}

 The real-time evolution operator, $e^{-  i \hat{H}  t } $, is a unitary operator if $\hat{H}$ is a Hermitian operator, thus  can be computed efficiently on unitary gate based quantum computers. The imaginary-time evolution operator, $e^{-   \hat{H}  \tau } $, can be obtained through analytic continuation that is described in Sec.~\ref{sec:analycontintro}. The spectral representation of   real-time evolution operator, $e^{-  i \hat{H}  t } $ is given by
 \begin{equation}
 e^{-  i \hat{H}  t }  = \sum_n | n \rangle \langle n | e^{- i \epsilon_nt},
 \end{equation}
 where $ | n \rangle $ and $\epsilon_n$ are the eigen-solutions of Hamiltonian operator $\hat{H}$: 
 \begin{equation}
 \hat{H}   | n \rangle = \epsilon_n | n \rangle .
 \end{equation}
 The Green's function operator of  $\hat{H}$,  $1/(E- \hat{H})$, is   related to  $e^{-  i \hat{H}  t } $ by Eq.(\ref{eq:realtoGreen}), hence the spectral representation of Green's function operator is given by
  \begin{equation} 
 \frac{1}{E -\hat{H}} = \sum_n  \frac{ | n \rangle \langle n |  }{E-  \epsilon_n } .
 \end{equation}
 Using Eq.(\ref{eq:realtoimagintegral}), the  transition probability distribution   in Eq.(\ref{eq:FPHtransitionprob}) can be related to quantum mechanical  Green's function by
   \begin{align}
 & P(x, \tau; x', 0)   \nonumber \\
 &= e^{ - \frac{\Phi (x)}{2}}       \sum_n    \langle x  | n \rangle \langle n | x' \rangle   e^{-  \epsilon_n  \tau}     e^{ \frac{\Phi (x')}{2}}    \nonumber \\
 &= e^{ - \frac{\Phi (x)}{2}}   \frac{i}{2\pi } \int_C \mathrm{d} E\     \langle x  |   \frac{  e^{- E \tau}  }{E-  \hat{H} }  | x' \rangle    e^{ \frac{\Phi (x')}{2}}   . 
 \end{align}

In this work, we will focus on two exactly solvable  process: (1) Brownian motion with a constant drift; and (2)  Ornstein-Uhlenbeck Process.

\subsection{Brownian motion with a constant drift} 
For Brownian motion with a constant drift, the drift coefficient function $\gamma (x)$ is given by constant,
\begin{equation}
\gamma (x) = \gamma,
\end{equation}
and
\begin{equation}
\Phi (x) = - \frac{ \gamma }{D} x,
\end{equation}
hence
\begin{align}
\hat{H}   =-    D \frac{\partial^2}{\partial x^2}   +   \frac{ \gamma^2}{4 D}   .
\end{align}
The energy spectral is continuous,  and eigen-solutions of $\hat{H}$ are given by plane waves,
\begin{equation}
\hat{H} e^{ i p x} = \epsilon_p e^{i p x},
\end{equation}
where  
\begin{equation}
 \epsilon_p   =  D p^2 +   \frac{ \gamma^2}{4 D}  .
\end{equation}
The energy spectral of Green's function of Hermitian Hamiltonian operator $\hat{H}$ is given by
\begin{align}
&   \langle x  |   \frac{  1 }{E-  \hat{H} }  | x' \rangle    \nonumber \\
&  = \int_{-\infty}^\infty \frac{d p}{2\pi} \frac{e^{i p x} e^{- i p x'}}{ E -  \epsilon_p  }    = -  \frac{i   e^{i \sqrt{\frac{1}{D} ( E - \frac{ \gamma^2}{4 D}  )} |x-x'|} }{ 2  \sqrt{ D ( E - \frac{ \gamma^2}{4 D}  )}}. 
\end{align}
The exact solution of transition probability distribution for Brownian motion with a constant drift is given by 
\begin{equation}
P(x, \tau ; x' , 0)   = \frac{1}{ \sqrt{4\pi D \tau} } e^{ - \frac{ (x - x' - \gamma \tau)^2}{4 D \tau}}.   \label{eq:ProbsolutionBM}
\end{equation}
Brownian motion with a constant drift  doesn't have non-trivial stationary solution, as time goes on, the transition probability distribution diffuse into zero completely.

\subsection{Ornstein-Uhlenbeck Process} 
For Ornstein-Uhlenbeck Process, the drift coefficient function $\gamma (x)$ is a linear function,
\begin{equation}
\gamma (x) = - \gamma x,
\end{equation}
and
\begin{equation}
\Phi (x) =  \frac{ \gamma }{2 D} x^2,
\end{equation}
hence
\begin{equation}
\hat{H}   =-    D \frac{\partial^2}{\partial x^2}   +   \frac{ \gamma^2}{4 D}  x^2 - \frac{\gamma}{2}  
\end{equation}
 is harmonic oscillator confining potential. The energy spectra of $\hat{H}$ for  Ornstein-Uhlenbeck Process are discrete valued. The eigen-solutions of $\hat{H}$ are given by
\begin{equation}
\hat{H} \phi_n (x) = \epsilon_n \phi_n (x) , 
\end{equation}
where 
\begin{equation}
\epsilon_n  = \gamma n  \ \   \mbox{with}  \ \ n =0,1, 2,  \cdots ,
\end{equation}
 and
\begin{equation}
\phi_n (x) =  \frac{1}{\sqrt{2^n n!} } \left ( \frac{\gamma}{2\pi D} \right )^{\frac{1}{4}} e^{ - \frac{\gamma  x^2}{4 D } } H_n ( \sqrt{ \frac{\gamma}{2D}} x ) .
\end{equation}
 The energy spectral of Green's function of Hermitian Hamiltonian operator $\hat{H}$ is given by
\begin{equation}
   \langle x  |   \frac{  1 }{E-  \hat{H} }  | x' \rangle   = \sum_n \frac{\phi_n (x)  \phi^*_n(x') }{E - \epsilon_n }   . 
\end{equation}
The exact solution of transition probability distribution for Ornstein-Uhlenbeck Process is given by 
\begin{equation}
P(x, \tau ; x' , 0)   =  \sqrt{\frac{\gamma}{2\pi D \left (1- e^{-2 \gamma \tau} \right )}} e^{ - \frac{\gamma (x-x' e^{-\gamma \tau})^2}{2 D \left (1- e^{-2 \gamma \tau} \right ) }}  . 
\end{equation}
As $\tau \rightarrow \infty$, the  transition probability distribution for  Ornstein-Uhlenbeck Process approaches the stationary limit where only the ground state with zero eigen-energy value survives,
\begin{align}
&P(x, \tau ; x' , 0)   \nonumber \\
&  \stackrel{\tau \rightarrow \infty}{\longrightarrow} e^{ - \frac{\Phi (x)}{2}}      \phi_0 (x)  \phi_0 (x')      e^{ \frac{\Phi (x')}{2}}     =  \sqrt{\frac{\gamma}{2\pi D }} e^{ - \frac{\gamma x^2}{2 D  }}  .  \label{eq:ProbsolutionOU}
\end{align}

\section{Bounds on cutoff dependence of double integrals in Eq.(\ref{eq:mainresult})}\label{sec:cutoff_dependence}
In this section, we present details of the cutoff dependence of the real-time integrals involved in reconstructing the ITE. 
To mirror the discussion in the main text, we provide two approaches: bounds derived from the direct approach, as well as bounds placed using the contour deformation perspective. 

\subsection{Direct approach bounds}\label{subsec:DAB}

In the direct approach, the relative error induced by truncating the integral to $\int_{-t_\mathrm{cut}}^{+t_\mathrm{cut}}\mathrm{d}t$ can be bounded by the following substitutions:
\begin{align}
&\left|\frac{\frac{i}{2 \pi} \int_{-t_\mathrm{cut}}^{+t_\mathrm{cut}} \mathrm{d}t\  \frac{e^{ic(t+i\tau)-iEt}}{t+i\tau}}{e^{-E\tau}} - 1\right| \nonumber \\
&=\left|\frac{i}{2 \pi e^{-E\tau}}\int_{|t|>t_\mathrm{cut}}\mathrm{d} t \ \frac{e^{ic(t+i\tau)-iEt}}{t+i\tau}\right|\nonumber \\
&=\left|\frac{e^{-c\tau}}{\pi e^{-E\tau}}\int_{t>t_\mathrm{cut}}\mathrm{d}t\ \frac{t \sin [(E-c)t] + \tau \cos[(E-c)t] }{t^2 + \tau^2}\right|. \nonumber
\end{align}
After using integration-by-parts, we find
\begin{align}
&\int_{t > t_\mathrm{cut}} \mathrm{d}t \ \frac{t \sin[(E-c)t]}{t^2 + \tau^2}  =\frac{-t \cos[(E-c)t]}{(E-c)(t^2 + \tau^2)} \Bigg|^{\infty}_{t_\mathrm{cut}} \nonumber \\
& + \int_{t_\mathrm{cut}}^\infty  \mathrm{d}t \frac{\cos[(E-c)t]}{E-c} \frac{d}{d t} \left( \frac{t}{t^2 + \tau^2}\right)  . \nonumber 
\end{align}
Also using $|\cos[(E-c)t] \leq 1$ and then assuming $t_\mathrm{cut} > \tau$, one has the bound:
\begin{equation}
\left| \int_{t > t_\mathrm{cut}} \mathrm{d}t \ \frac{t \sin[(E-c)t]}{t^2 + \tau^2}\right| \leq \frac{2t_\mathrm{cut}}{(E-c) (t_\mathrm{cut}^2 + \tau^2)}.
\end{equation}
Applying the same method to the $\mathrm{cos}$ term, one derives the desired bound:
\begin{equation}
\left|\frac{\frac{i}{2 \pi} \int_{-t_\mathrm{cut}}^{+t_\mathrm{cut}} \mathrm{d}t\  \frac{e^{ic(t+i\tau)-iEt}}{t+i\tau}}{e^{-E\tau}} - 1\right| \leq \frac{2}{\pi}\frac{t_\mathrm{cut} + \tau}{t_\mathrm{cut}^2 + \tau^2} \frac{e^{(E-c)\tau}}{E-c}.
\end{equation}

\subsection{Contour Deformation bounds}\label{subsec:CDB}

Placing a cutoff $t_\mathrm{cut}$ on the $t$ integrals defining the $\hat{I}_{1,2} (\epsilon)$ operators shown in Eqs~(\Ref{eq:I12functions},\ref{eq:mainresult}), 
\begin{align}
\hat{I}_1 (\epsilon,t_\mathrm{cut}) & = \int_0^{t_\mathrm{cut}} \mathrm{d} t\  e^{i (c+i \epsilon -  \hat{H} )  t}  =  \frac{i - i e^{    i ( c  - \hat{H}  ) t_\mathrm{cut}} e^{-\epsilon t_\mathrm{cut}} }{c+  i \epsilon - \hat{H}}  , \nonumber \\
\hat{I}_2 (\epsilon,t_\mathrm{cut}) & =  \int_0^{t_{cut}} \mathrm{d} t\  e^{i (\epsilon + i a- \hat{H}) t}  =   \frac{i -  ie^{i (\epsilon   - \hat{H}) t_\mathrm{cut} } e^{ -  a   t_\mathrm{cut} } }{ \epsilon + i a - \hat{H}}    ,
\end{align} 
from which we derive the asymptotic estimates: 
\begin{align}
  \hat{I}_1 (\epsilon,t_\mathrm{cut}) &   \simeq   \hat{I}_1 (\epsilon) +  \mathcal{O} ( e^{-\epsilon t_\mathrm{cut}} ) , \nonumber \\
 \hat{I}_2 (\epsilon,t_\mathrm{cut}) & \simeq  \hat{I}_2 (\epsilon)   + \mathcal{O} (  e^{ -  a   t_\mathrm{cut} } )  . \label{eq:I12bound}
\end{align} 
Numerically  we need to make sure that $\epsilon t_{cut} $ and $ a t_{cut} $ are sufficient large. Let's introduce another scale $\Lambda$, so that $e^{- \Lambda} $ is numerically acceptable for a specific application, hence, we need to make sure that
\begin{equation}
\epsilon > \frac{\Lambda}{t_\mathrm{cut}}, \ \ \ \  a > \frac{\Lambda}{t_\mathrm{cut}}.
\end{equation}

As shown in Eq.(\ref{eq:I12bound}), the convergence of $ \hat{I}_2 (\epsilon,t_\mathrm{cut}) $ toward $ \hat{I}_2 (\epsilon)  $ doesn't depend on the $\epsilon$, hence the double integrals involving $ I_2 (\epsilon,t_\mathrm{cut}) $,
 \begin{equation}
    \frac{1}{2\pi }   \int_{c }^{  \infty }   \mathrm{d} \epsilon \  e^{- (\epsilon +  i a) \tau}  \hat{I}_2 (\epsilon, t_{cut}) ,
\end{equation}
is soundly defined, and the bound on double integrals involving $ \hat{I}_2 (\epsilon,t_\mathrm{cut}) $ is given by
 \begin{align}
  &  \frac{1}{2\pi }   \int_{c }^{  \infty }   \mathrm{d} \epsilon \  e^{- (\epsilon +  i a) \tau}    \hat{I}_2 (\epsilon,t_{cut}) \nonumber \\
  &  \simeq    \frac{1}{2\pi }   \int_{c }^{  \infty }   \mathrm{d} \epsilon \  e^{- (\epsilon +  i a) \tau}   \hat{I}_2 (\epsilon)  +  \mathcal{O} (  e^{ -  a   t_{cut} } ).
\end{align}

Since the convergence of  $ \hat{I}_1 (\epsilon, t_\mathrm{cut}) $ toward $ \hat{I}_1 (\epsilon)  $ depends on $\epsilon$,  we need to be careful on the analysis of the double integrals involving $ \hat{I}_1 (\epsilon, t_\mathrm{cut}) $. Starting with  a numerically well defined quantity,
 \begin{equation}
   \frac{i}{2\pi }    \int_{\frac{\Lambda}{t_{cut}} }^{ a}   \mathrm{d} \epsilon\  e^{- (c+i \epsilon) \tau}  \hat{I}_1 (\epsilon, t_\mathrm{cut})     , \label{eq:I1tcut}
\end{equation}
where we need to make sure that $\frac{\Lambda}{t_\mathrm{cut}} \rightarrow 0$ and $t_\mathrm{cut} \gg \Lambda$, then   Eq.(\ref{eq:I1tcut})  can be rearranged to
 \begin{align}
 &  \frac{i}{2\pi }    \int_{\frac{\Lambda}{t_{cut}} }^{ a}   d \epsilon e^{- (c+i \epsilon) \tau}  \hat{I}_1 (\epsilon, t_{cut})     \nonumber \\
 & = \frac{i}{2\pi }    \int_{ 0}^{ a}   d \epsilon e^{- (c+i \epsilon) \tau}  \hat{I}_1 (\epsilon)    \nonumber \\
 & +  \frac{i}{2\pi }    \int_{\frac{\Lambda}{t_{cut}} }^{ a}   d \epsilon e^{- (c+i \epsilon) \tau}  \left [ \hat{I}_1 (\epsilon, t_{cut})  -  \hat{I}_1 (\epsilon)  \right ]  \nonumber \\
 & -  \frac{i}{2\pi }    \int_{ 0}^{ \frac{\Lambda}{t_{cut}} }   d \epsilon e^{- (c+i \epsilon) \tau}  \hat{I}_1 (\epsilon)    . \label{eq:I1cutrearrange}
 \end{align}
The second term and the third term in right-hand-side of Eq.(\ref{eq:I1cutrearrange}) are bound by
 \begin{equation}
     \frac{i}{2\pi }    \int_{\frac{\Lambda}{t_{cut}} }^{ a}   \mathrm{d} \epsilon\  e^{- (c+i \epsilon) \tau}  \left [\hat{I}_1 (\epsilon, t_{cut}) -  \hat{I}_1 (\epsilon)  \right ]   \approx       \mathcal{O} (   e^{ -  \Lambda } ) 
 \end{equation}
and
 \begin{equation}
     \frac{i}{2\pi }    \int_{ 0}^{ \frac{\Lambda}{t_{cut}} }   \mathrm{d} \epsilon\  e^{- (c+i \epsilon) \tau}  \hat{I}_1 (\epsilon)   \approx \mathcal{O} (   \frac{\Lambda}{t_{cut}}    ) ,
 \end{equation}
therefore,  the bound on double integrals involving $ \hat{I}_1 (\epsilon,t_\mathrm{cut}) $ is given by 
  \begin{align}
 &  \frac{i}{2\pi }    \int_{\frac{\Lambda}{t_\mathrm{cut}} }^{ a}   \mathrm{d} \epsilon\  e^{- (c+i \epsilon) \tau}  \hat{I}_1 (\epsilon,t_{cut})      \nonumber \\
 &   \simeq       \frac{i}{2\pi }    \int_{ 0}^{ a}   \mathrm{d} \epsilon\  e^{- (c+i \epsilon) \tau}  \hat{I}_1 (\epsilon) +   \mathcal{O} (   e^{ -  \Lambda } )    +   \mathcal{O} (   \frac{\Lambda}{t_\mathrm{cut}}    ).
 \end{align}

\bibliography{ALL-REF.bib}

\end{document}